\documentclass[pre,twocolumn,showpacs,floatfix]{revtex4}
\usepackage{dcolumn}% Align table columns on decimal point
\usepackage{bm}% bold math
\usepackage{epsfig}
\RequirePackage{times}
\RequirePackage[dvips]{hyperref}
\begin{document}
\title{ On the nonlinear dynamics of topological solitons in DNA}
\author{L. V. Yakushevich}
\email{ykushev@icb.psn.ru}
\affiliation{Institute of Cell Biophysics, Russian Academy of Sciences,
142290 Pushchino, Russia}
\author{A. V. Savin}
\email{asavin@center.chph.ras.ru}
\affiliation{
Institute for Physics and Technology, ul. Prechistenka 13/7,
119034 Moscow, Russia}
\author{L. I. Manevitch}
\email{lmanev@center.chph.ras.ru}
\affiliation{N. N. Semenov Institute of Chemical Physics,
Russian Academy of Sciences,
ul. Kosygina 4, 117977 Moscow, Russia}

%\date{\today}
\date{April 30, 2002}

\begin{abstract}
Dynamics of topological solitons describing open states in the DNA double helix
are studied in the frameworks of the model which takes into account
asymmetry of the helix. It is shown that three types of topological solitons
can occur in the DNA double chain. Interaction between the solitons,
their interactions with  the chain inhomogeneities  and stability of the solitons
with respect to thermal oscillations are investigated.
\end{abstract}



\maketitle

\section{Introduction}
It is widely accepted now that the DNA molecule has a rather
moveable internal structure, and that the internal DNA mobility
plays an important role in functioning the molecule. The thermal
bath where the DNA molecule is usually immersed, collisions with
the molecules of the solution which surrounds DNA, local
interactions with proteins, drugs or some other ligands lead to
activation of different types of internal motions. Small
oscillations of individual atoms near equilibrium positions,
rotational, transverse and longitudinal displacements of atomic
groups (phosphate groups, sugars and bases), motions of the double
chain fragments having several base pairs lengths, local unwinding
of the double helix, transitions of DNA fragments from one
conformational form to another, for example, from A-form to B-form
and so on, are only some of them. A more detailed list of internal
motions and of their dynamical characteristics can be found in the
works of Fritzsche \cite{p1}, Keepers and co-authors \cite{p2},
McClure \cite{p3}, McCommon and co-authors \cite{p4}, Yakushevich
\cite{p5,p6}).

Different approaches to the modeling of the internal DNA mobility
are known. One of them has been developed by Prohofsky and co-authors
\cite{p7,p8,p9,p10}, who considered DNA as a lattice and took
into account the motions of all atoms (except of hydrogen atoms)
in the lattice cell. Their approach was limited, however, by harmonic
approximation, and this limitation did not
permit them to model large amplitude internal motions such as, for
example, local unwinding of the double helix. Another approach,
based on the methods of molecular dynamics and proposed firstly
by Levitt \cite{p11} and Tidor and co-authors \cite{p12}, is known
now as one of the most powerful tools of investigation of the
internal DNA mobility \cite{p13}. This approach is not limited by
harmonic approximation and therefore it can be used to study internal
motions of both large and small amplitudes. The approach has, however,
one essential deficiency: because of the limited possibilities of modern
computers it can not be used to study long DNA fragments, and therefore
it does not suitable for studies of the processes of propagation
of local structural distortions along the molecule.

In this paper, to investigate the internal DNA mobility, we use the approach
developed in a series of works
\cite{p14,p15,p16,p17,p18,p19,p20,p21,p22,p23,p24,p25}.
Peculiarity of the
approach is that it uses rather simple models of the internal DNA
dynamics, which take into account only one or a few types of the DNA
internal motions. This simplification gives an opportunity to find
analytical solutions of corresponding dynamical equations imitating
both small and large amplitude internal motions.
And one more merit of the approach is that it gives a possibility to study
the internal dynamics of long DNA fragments. Three works in the series are
of most interest.

The first one has been done by Englander and co-authors \cite{p14}
who studied the dynamics of DNA open states. Their model took into
account only rotational motions of nitrous bases, which as it was
suggested, made the main contribution to formation of the open
states. Another paper belonged to Peyrard and Bishop \cite{p22},
who studied the process of DNA denaturation. Suggesting that the
stretching of the hydrogen bonds in pairs made the most
contribution into the process, they created a simplified model
where only transverse motions of bases along the direction of the
hydrogen bonds where taken into account. The third important paper
was published by Muto and co-authors \cite{p21}. These authors
suggested that two types of internal motions were made the main
contribution to DNA denaturation process: transverse motions along
the hydrogen bond direction and longitudinal motions along the
backbone direction. Their model consisted of two polynucleotide
strands linked together through the hydrogen bonds described by a
Lennard-Jones potential, and the phosphodiester bridges in the
backbone were described by an anharmonic Toda potential.

%-------------------------------------------------------------------------
\begin{figure}[htb]
\centering\epsfig{file=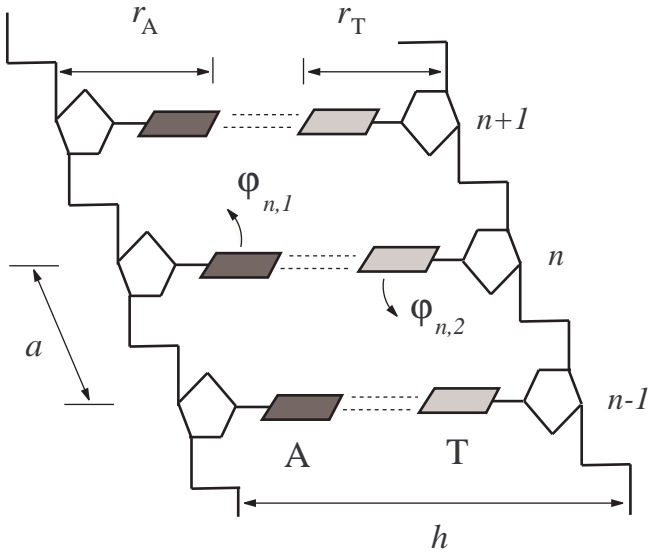,width=0.99\linewidth}
\caption{\label{fig1}\protect
        Fragment of the DNA double chain consisting of 3 AT base pairs.
        Longitudinal pitch of the helix $a=3.4$ \AA; \
        transverse pitch $h=16.15$ \AA .
        }
 \end{figure}
%-------------------------------------------------------------------------

Further
development of the
approach was limited for several years by small improvements of the models and their
combinations, and only involving numerical methods of simulation
of the internal DNA dynamics gave a new impulse and interesting
possibilities which have been realized in the works of Van Zandt \cite{p26},
Techera and co-authors \cite{p27},  Salerno \cite{p25},
Barbi  \cite{p28,p29}, and Campa \cite{p30}. Just these methods
permitted not only to study a possibility of appearance of large
amplitude localized distortions in the DNA structure, but also to
investigate their stability, the influence of thermal noise, the
interactions between the distortions, the propagation of them along
the homogeneous and inhomogeneous DNA.
%------------------------------------------------------------------
\begin{table}[t]
\caption{\label{tab1} The values of the parameters $m_\alpha$,
$r_\alpha$, $I_\alpha=m_\alpha r_\alpha^2$, for all possible bases
$\alpha$ ($m_p=1.67343\cdot 10^{-27}$ kg is the proton mass).}
\begin{ruledtabular}
\begin{tabular}{lccr}
 $\alpha$ & $m_\alpha$ $(m_p)$ & $r_\alpha$ (\AA ) &
 $I_\alpha$ ($\times 10^{-47}$ m$^2$kg) \\
 \hline
 A & 135.13 & 5.8 & 7607.03 \\
 T & 126.11 & 4.8 & 4862.28 \\
 G & 151.14 & 5.7 & 8217.44 \\
 C & 111.10 & 4.7 & 4106.93 \\
\end{tabular}
\end{ruledtabular}
\end{table}
%------------------------------------------------------------------

In all these works, however, the
asymmetry of the base pairs was neglected. That is both bases
in a pair were modeled as identical structural elements with the
same characteristics (masses, moments of inertia and so on). But
even in the case of homogeneous (synthetic) DNA the asymmetry exists.
Indeed, if, for example,  one of the polynucleotide chains consists of
only adenines, the other chain should consist of thymines, and this
homogeneous model is substantially asymmetrical. Just this type of asymmetrical
model is studied in this work. To simplify calculations, we consider
only rotational motions of nitrous bases around the sugar-phosphate chains
in the plane perpendicular to the main axis of the double chain. We find
solitary wave solutions describing open states in the double helix.
We classify the solitons, investigate stability of the solitons with respect
to thermal oscillations, interactions between the solitons,
interaction of the solitons with inhomogeneities of the chain.
To solve all these problems, we use numerical-variation methods efficiency
of which was proved in the works  \cite{p31,p32,p33,p34,p35,p36},
devoted to the analysis of nonlinear dynamics of molecular chains and
polymer crystals.\\

\section{Discrete model of the DNA double helix}

Let us consider B-form of the DNA molecule, the fragment of
which is presented in Fig. \ref{fig1}. The lines in the figure
correspond to the skeleton of the double helix, black and grey rectangles
correspond to bases in pairs (AT and GC). Let us focus our attention
on the rotational motions of bases around the sugar-phosphate chains
in the plane perpendicular to the helix axis. Below we shall call the
chain placed on the left by the first chain, and the right chain --
by the second one. Positive directions of the rotations of the bases
for each of the chains are shown in Fig. \ref{fig1}.

Let us consider the plane DNA
model where the chains of the macromolecule form two parallel straight
lines placed at a distance $h$ from each other, and the bases can make
only rotation motions around their own chain, being all the time
perpendicular to it. Let us suggest that $\varphi_{n,1}$ is the angular
displacement of the $n$-th base of the first chain, and $\varphi_{n,2}$
is the angular displacement of the $n$-th base of the second chain.
Then the Hamiltonian of the double chain takes the form
\begin{widetext}
\begin{equation}
H=\sum_n\left\{
\frac12 I_{n,1}\dot{\varphi}_{n,1}^2
                +\frac12 I_{n,2}\dot{\varphi}_{n,2}^2
   +\epsilon_{n,1}\sin^2\frac{\varphi_{n+1,1}-\varphi_{n,1}}{2}+
           \epsilon_{n,2}\sin^2\frac{\varphi_{n+1,2}-\varphi_{n,2}}{2}
   + V_{\alpha\beta}(\varphi_{n,1},\varphi_{n,2})\right\}~.
               \label{f1}
\end{equation}
\end{widetext}
The first two terms of Hamiltonian (\ref{f1}) correspond to the
kinetic energy of the $n$-th base pair. Here $I_{n,1}$ is the moment
of inertia of the $n$-th base of the first chain; $I_{n,2}$ is the moment
of inertia of the $n$-th base of the second chain, point denotes
differentiation in time $t$. For the base pair $\alpha\beta$
$(\alpha\beta=AT,~TA,~CG,~GC)$ the moment of inertia is equal to
$I_{n,1}=m_\alpha r_\alpha^2$, $I_{n,2}=m_\beta r_\beta^2$.
The value of the base mass $m_\alpha$, the length $r_\alpha$ and
corresponding moment of inertia $I_\alpha=m_\alpha r_\alpha^2$ for
all possible base pairs are presented in the Table \ref{tab1}.

The third and the fourth terms in Hamiltonian (\ref{f1}) describe
interaction of the neighboring bases along each of the
macromolecule chains. Parameter $\epsilon_{n,i}$ characterizes the
energy of interaction of the $n$-th base with the $(n+1)$-th base
of the $i$-th chain $(i=1,2)$. The value of the parameter is
unknown. But if we take into account that angular displacement of
one base is accompanied not only by overcoming the barrier due to
the stacking interaction, but also by substantial deformation of
the dihedral and valence angles,we can suggest that the energy of
the displacement $\epsilon_{n,i}$ should be wittingly more than
the stacking $40\div 60$ kJ/mol \cite{p37}, and it should weakly
depend on the type of the base. This gives us a possibility to
suggest later on that
$\epsilon_{n,1}\equiv\epsilon_{n,2}\equiv\epsilon> 60$ kJ/mol.

The fifth term in Hamiltonian (\ref{f1}) corresponds to the energy of
interaction between conjugated bases of different chains.
Here index $\alpha\beta =$AT, TA, GC, CG determines the type of the
base pair. It is convenient to model the energy of interaction
of conjugated pairs by the potential
\begin{equation}
V_{\alpha\beta}(\varphi_{n,1},\varphi_{n,2})=\frac12 K_{\alpha\beta}
                |{\bf R}_n-{\bf R}_n^\circ|^2~,
\label{f2}
\end{equation}
where {\bf R}$_n$ is the vector connecting the end of the base $(n,1)$
with the end of the base $(n,2)$, {\bf R}$_n^\circ$ is the value
of the vector for the ground state of the chain $\varphi_{n,1}\equiv 0$,
$\varphi_{n,2}\equiv 0$. Potential (\ref{f2}) can be written in a
more simple form

\begin{widetext}
\begin{equation}
V_{\alpha\beta}(\varphi_{n,1},\varphi_{n,2})=K_{\alpha\beta}
\{r_\alpha(r_\alpha+r_\beta)(1-\cos\varphi_{n,1})+
 r_\beta(r_\alpha+r_\beta)(1-\cos\varphi_{n,2})-
 r_\alpha r_\beta[1-\cos(\varphi_{n,1}-\varphi_{n,2})]\}~.
\label{f3}
\end{equation}
\end{widetext}

The rigidity of interaction $K_{\alpha\beta}$ can be estimated from
the energy of interaction
$$
e_{\alpha\beta}=\frac12\left[V_{\alpha\beta}(0,\frac{\pi}{2})+
V_{\alpha\beta}(\frac{\pi}{2},0)\right]
=\frac12K_{\alpha\beta}[r_\alpha^2+r_\beta^2]~.
$$
The pair AT (TA) is stabilized by two hydrogen bonds (they are shown
in Fig. \ref{fig1} by dotted lines), and the pair CG (GC) --
by three hydrogen bonds. Therefore we suggest later on that
$e_{AT}=e_{TA}=2e_{CG}/3=2e_{GC}/3=e$.

For the value of the energy of interaction of the bases in AT base
pair we can take the double energy of hydrogen bond $e=40$ kJ/mol.
Then the rigidity of the bond between the bases is equal to
\begin{eqnarray}
K_{AT}=K_{TA}=\frac23K_{GC}=\frac23K_{CG}=\nonumber \\
K=\frac{2e}{r_\alpha^2+r_\beta^2}
=0.234 \mbox{ N/m}.  \label{f4}
\end{eqnarray}
On the other hand, the value of the parameter can be estimated
from the frequency spectrum of low amplitude oscillations of the chain.
We shall obtain it in the next section.
%-------------------------------------------------------------------------
\begin{figure*}[tbh]
\centering\epsfig{file=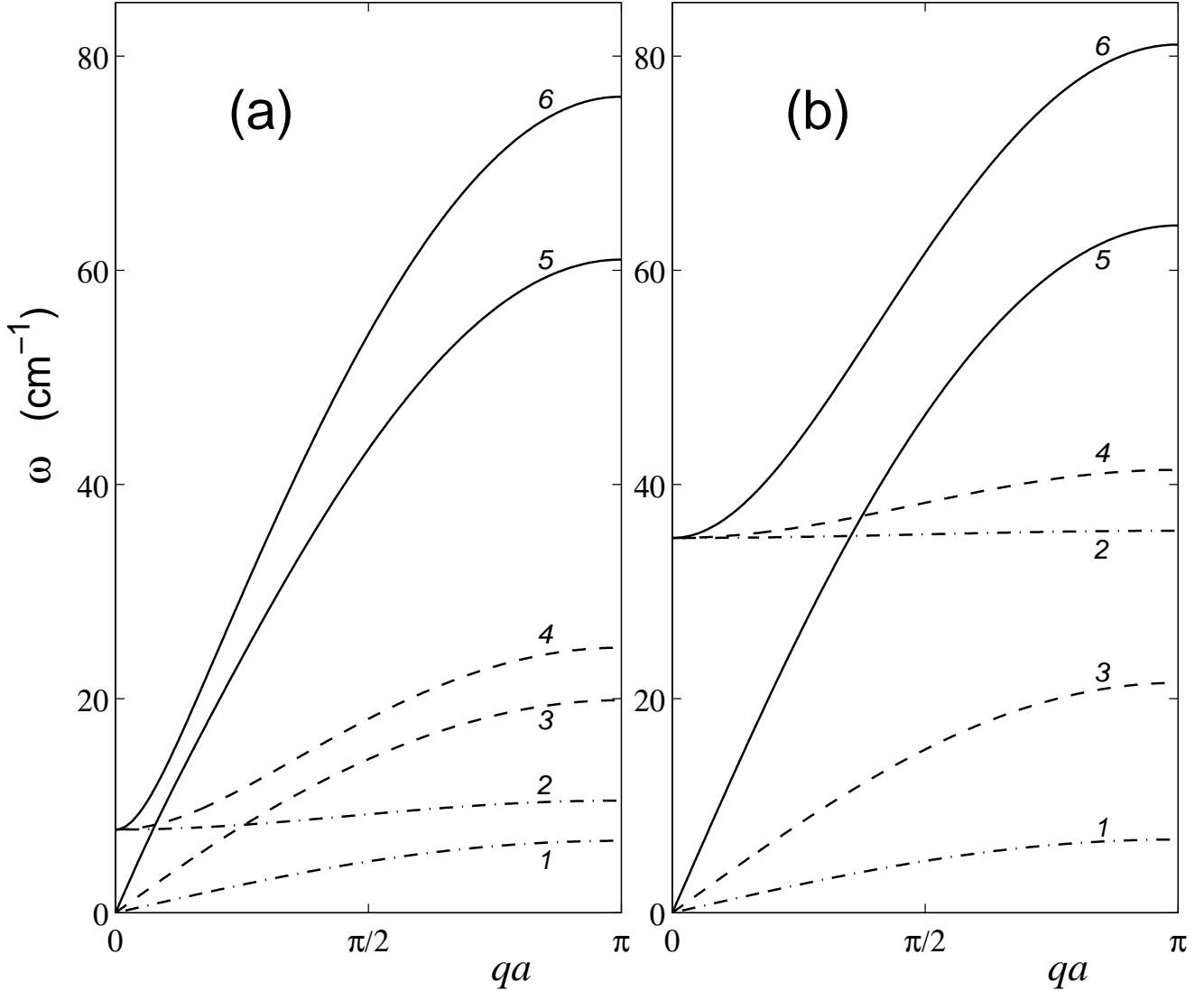,width=0.99\linewidth}
\caption{\label{fig2}\protect
        Acoustic $\omega=\omega_a(q)$ (curve 1, 3, 5) and optical
        $\omega=\omega_o(q)$ (curve 2, 4, 6) branches of the dispersion curve
        for homogeneous chain ($\alpha\beta=$AT, $\epsilon=$60, 600, 6000 kJ/mol)
        for chain with $K=0.234$ N/m (a) and $K=4.744$ N/m (b).}
 \end{figure*}
%-------------------------------------------------------------------------

\section{Dispersion equation}

The system of equations of motion, which corresponds to macromolecule
Hamiltonian (\ref{f1}), takes the form
\begin{eqnarray}
I_{n,1}\ddot{\varphi}_{n,1} &=& -\frac{\partial H}{\partial\varphi_{n,1}}~;
\nonumber \\
I_{n,2}\ddot{\varphi}_{n,2} &=& -\frac{\partial H}{\partial\varphi_{n,2}}~,
\label{f5} \\
     n &=& 0,\pm1,\pm2,...~~. \nonumber
\end{eqnarray}
Let us consider homogeneous macromolecule where only one type of
base pairs exists $\alpha\beta$
($I_{n,1}\equiv I_\alpha$, $I_{n,2}\equiv I_\beta$).

Insert small amplitude plane wave
$$
(\varphi_{n,1}(t),\varphi_{n,2}(t))=
(\varphi_1,\varphi_2)Ae^{i(qan-\omega t)}~,
$$
into the system of equations (\ref{f5}). Here $(\varphi_1,\varphi_2)$
is a normalized to 1 two dimension vector,
$A\ll \pi$ is the amplitude, $q\in [0,\pi /a]$ is the wave number.
It is easy to show that in the linear approximation the frequency
$\omega$ should satisfy the dispersion equation
\begin{equation}
\omega^4+B\omega^2+C=0~, \label{f6}
\end{equation}
where
\begin{eqnarray}
 B&=&[K_{\alpha\beta}(I_\alpha r_\beta^2+ I_\beta r_\alpha^2)
 +4\kappa (I_\alpha +I_\beta )\sin^2\frac{qa}{2})]/I_{\alpha}I_\beta~;
  \nonumber \\
 C&=&[4\kappa K_{\alpha\beta}(r_\alpha^2+r_\beta^2)\sin^2\frac{qa}{2}
 +16\kappa^2\sin^4\frac{qa}{2}]/I_{\alpha}I_\beta~, \nonumber
\end{eqnarray}
$\kappa=\epsilon/2$ is the rigidity of the interaction of neighboring
bases along the chain.

Dispersion curve (\ref{f6}) has two branches:
\begin{eqnarray}
\omega_a(q)&=&[(B-\sqrt{B^2-4C})/2]^{1/2}~; \nonumber \\
\omega_o(q)&=&[(B+\sqrt{B^2-4C})/2]^{1/2}~. \nonumber
\end{eqnarray}
The upper curve $\omega=\omega_o(q)$ corresponds to optical phonons,
the lower curve $\omega=\omega_a(q)$ corresponds to acoustic phonons
in the chain.

The frequency $\omega_a(q)$ tends to zero as $q\rightarrow 0$.
Let us determine the velocity of acoustic phonons as
$$
v_0=\lim_{q\rightarrow 0}\frac{\omega(q)}{q}
=a\sqrt{\frac{\kappa (r_\alpha^2+r_\beta^2 )}
{I_\alpha r_\beta^2+I_\beta r_\alpha^2}}~.
$$
The dependence of the sound velocity $v_0$ in the homogeneous molecule
$\alpha\beta=$AT (GC) on the energy of rotation
$\epsilon$ is presented in the Table \ref{tab2}.

According to different estimations \cite{p38,p39,p40} the velocity
of sound in DNA is on the interval from 1890 m/s till 3500~m/s.
From the Table \ref{tab2} it is clear, that among three typical
values $\epsilon=60$, 600, 6000~kJ/mol the value
$\epsilon=6000$~kJ/mol is the best one. Just this value will be
used in the numerical investigations of the dynamics of
topological solitons.

%-------------------------------------------------------------------------
\begin{table}[b]
\caption{\label{tab2}
Dependence of sound velocity $v_0$ (m/s) on the value of the parameter
$\epsilon$ for homogeneous $\alpha\beta=$AT (GC) chain}
\begin{ruledtabular}
\begin{tabular}{lccr}
$\epsilon$ (kJ/mol) &  60  & 600 & 6000 \\
\hline
AT & 219.47 & 694.02 & 2194.7 \\
GC & 223.38 & 706.39 & 2233.4 \\
\end{tabular}
\end{ruledtabular}
\end{table}
%-------------------------------------------------------------------------

The lowest value of the optical frequency is
\begin{equation}
\omega_o(0)=\sqrt{K_{\alpha\beta}(I_\alpha r_\beta^2+I_\beta
r_\alpha^2)/I_\alpha I_\beta}~. \label{f7}
\end{equation}
According to \cite{p41} $\omega_o(0)=35$ cm$^{-1}$, therefore from
(\ref{f7}) we have
\begin{equation}
K_{AT}=K=4.744\mbox{ N/m},~~K_{CG}=\frac32K=7.117\mbox{ N/m}. \label{f8}
\end{equation}
This estimation of the value of the rigidity differs from that
obtained in (\ref{f4}). The use of the value given in (\ref{f4})
gives substantially lower value of the frequency
$\omega_o(0)=7.77$ cm$^{-1}$. Thus we have the following estimation of
the value of the parameter $K$: $0.234\le K\le 4.744$ N/m. The view
of the dispersion curves for homogeneous chain ($\alpha\beta=AT$) with
different values of the parameters is presented in Fig. \ref{fig2}.

For numerical investigation of the soliton dynamics we shall take
intermediate value $K=0.8714$ N/m which corresponds to the frequency
$\omega_o(0)=15$ cm$^{-1}$, and energy of interaction $e_{AT}=149$ kJ/mol.

\section{Numerical method of finding solitary wave solutions}
Complexity of the system of equations of motions (\ref{f5}) does
not permit us to carry out analytical investigation. Therefore, we shall study
it numerically and use variation technique, proposed in \cite{p32}, to find
soliton like solutions.

Let us consider homogeneous DNA molecule (for all $n$ $I_{n,1}=I_\alpha$,
$I_{n,2}=I_\beta$, where $\alpha\beta=$ AT (TA, CG, GC)).
We shall find the solution of system (\ref{f5}) in the
form of a wave with smooth constant profile. For the purpose,
let us suggest that $\varphi_{n,1}(t)=\varphi_1(\xi)$,
$\varphi_{n,2}(t)=\varphi_2(\xi)$,
where the wave variable $\xi=na-vt$, and $v$ is the velocity of the wave.

Let us assume, that the functions $\varphi_1$ and $\varphi_2$ smoothly
depend on $\xi$. Then the time second derivatives can be substituted
for discreet derivatives
\begin{equation}
\frac{d^2\varphi_{n,i}}{dt^2}=v^2\frac{\partial\varphi_i}{\partial\xi^2}
=v^2(\varphi_{n+1,i}-2\varphi_{n,i}+\varphi_{n-1,i})/a^2,
\label{f9}
\end{equation}
$i=1,2$. Using these relations, we can write the equations of motions
(\ref{f5}) in the form
\begin{equation}
L_{\varphi_{n,1}}=0,~~L_{\varphi_{n,2}}=0,~~n=0,\pm 1,\pm 2,...~.
\label{f10}
\end{equation}
Here the functional
\begin{eqnarray}
L=\sum_n\{\frac{v^2}{2a^2}[I_\alpha (\varphi_{n+1,1}-\varphi_{n,1})^2
+I_\beta (\varphi_{n+1,2}-\varphi_{n,2})^2] - \nonumber \\
-\epsilon(\sin^2\frac{\varphi_{n+1,1}-\varphi_{n,1}}{2}
+\sin^2\frac{\varphi_{n+1,2}-\varphi_{n,2}}{2})\nonumber \\
-V_{\alpha\beta}(\varphi_{n,1},\varphi_{n,2})\} \nonumber
\end{eqnarray}
is a discreet version of the Lagrangian
\begin{eqnarray}
{\cal L}=\sum_n
[\frac12 I_{n,1}\dot{\varphi}_{n,1}^2
                +\frac12 I_{n,2}\dot{\varphi}_{n,2}^2-  \nonumber\\
  -\epsilon (\sin^2\frac{\varphi_{n+1,1}-\varphi_{n,1}}{2}+
           \sin^2\frac{\varphi_{n+1,2}-\varphi_{n,2}}{2})\nonumber \\
           -V_{\alpha\beta}(\varphi_{n,1},\varphi_{n,2})]~,
  \nonumber
\end{eqnarray}
which corresponds to the system of equations of motion (\ref{f5}).

For further analysis it is convenient to write the functional $L$
in the dimensionless form
\begin{eqnarray}
\bar{L}=2L/K(r_A^2+r_T^2)=
\nonumber \\
\sum_n[c_\alpha (\varphi_{n+1,1}-\varphi_{n,1})^2
+c_\beta (\varphi_{n+1,2}-\varphi_{n,2})^2\nonumber \\
-g(\sin^2\frac{\varphi_{n+1,1}-\varphi_{n,1}}{2}
+\sin^2\frac{\varphi_{n+1,2}-\varphi_{n,2}}{2})\label{f11} \\
-U_{\alpha\beta}(\varphi_{n,1},\varphi_{n,2})], \nonumber
\end{eqnarray}
where the dimensionless coefficients
$$
c_\alpha=\frac{v^2I_\alpha}{Ka^2(r_A^2+r_T^2)},~~
c_\beta=\frac{v^2I_\beta}{Ka^2(r_A^2+r_T^2)},
$$
parameter of cooperativity
\begin{equation}
g=2\epsilon/K(r_A^2+r_T^2), \label{f12}
\end{equation}
dimensionless potential $U_{\alpha\beta}(\varphi_{n,1},\varphi_{n,2})=
2V_{\alpha\beta}(\varphi_{n,1},\varphi_{n,2})/K(r_A^2+r_T^2)$.

Soliton solution of the system (\ref{f10}) can be found numerically
as a solution of the problem on conditional minimum
\begin{eqnarray}
-\bar{L}\rightarrow \min_{\varphi_{2,i},...,\varphi_{N-1,i},~i=1,2}:
 \label{f13}\\
\varphi_{1,1}=\varphi_{-\infty,1},~
\varphi_{1,2}=\varphi_{-\infty,2},\label{f14}\\
\varphi_{N,1}=\varphi_{\infty,1},~
\varphi_{N,2}=\varphi_{\infty,2}~. \label{f15}
\end{eqnarray}
Boundary conditions (\ref{f14}), (\ref{f15}) for the problem
(\ref{f13}) determine the type of the soliton solution.
We should take a rather large number $N$, in order that the form
of the solution of the problem might not depend on its value.
For the purpose it is enough to take $N$ ten times larger than
the width of the soliton.

The soliton solution of the problem (\ref{f13}) can be characterized
by the topological charge {\bf q}=$(q_1,q_2)$, where
$q_i=(\varphi_{\infty,i}-\varphi_{-\infty,i})/2\pi$, $i=1,2$, is
an integer $(q_i=0,\pm 1,\pm 2,...)$. To find soliton solution
with topological charge {\bf q}, it is necessary to solve the problem
on minimum (\ref{f13}) with boundary conditions
$$
\varphi_{-\infty,1}=\varphi_{-\infty,2}=0,~~
\varphi_{\infty,1}=2\pi q_1,~~\varphi_{\infty,2}=2\pi q_2~.
$$
This problem was solved by the method of conjugated gradient.
The value $N=2000$ was taken, and the initial point
$$
\varphi_{n,i}=[1+\tanh \mu (n-N/2)]\pi q_i~,~~i=1,2~,
$$
was used. Here $\mu$ is a changeable parameter.

Soliton solution (solution in the form of a solitary wave)
$\{\varphi_{n,1}^\circ,\varphi_{n,2}^\circ\}_{n=1}^N$ corresponds
to topological soliton with the energy
$E=K(r_A^2+r_T^2)\bar{E}/2$, where dimensionless energy
\begin{eqnarray}
\bar{E}=\sum_{n=1}^{N-1}[c_\alpha (\varphi_{n+1,1}-\varphi_{n,1})^2
+c_\beta (\varphi_{n+1,2}-\varphi_{n,2})^2 + \nonumber \\
+g(\sin^2\frac{\varphi_{n+1,1}-\varphi_{n,1}}{2}
+\sin^2\frac{\varphi_{n+1,2}-\varphi_{n,2}}{2})\nonumber \\
+U_{\alpha\beta}(\varphi_{n,1}\varphi_{n,2})] \nonumber
\end{eqnarray}
and with the diameter
$$
D=1+2\sqrt{(n-\bar{n})^2p_n}~,
$$
where the point
$$
\bar{n}=\sum_{n=1}^N np_n
$$
determines the position of the soliton center, and the formula
\begin{widetext}
\begin{eqnarray}
p_n=\bar{E}_n/\bar{E}=
\left\{\frac{1}{4}\left[c_\alpha (\varphi_{n+1,1}-\varphi_{n-1,1})^2
+c_\beta (\varphi_{n+1,2}-\varphi_{n-1,2})^2\right]
+\frac12 g\left[
 \sin^2\frac{\varphi_{n+1,1}-\varphi_{n,1}}{2} \right. \right. \nonumber \\
\left. \left.
+\sin^2\frac{\varphi_{n+1,2}-\varphi_{n,2}}{2}
+\sin^2\frac{\varphi_{n,1}-\varphi_{n-1,1}}{2}
+\sin^2\frac{\varphi_{n,2}-\varphi_{n-1,2}}{2}\right]
+U_{\alpha\beta}(\varphi_{n,1}\varphi_{n,2})\right\}/\bar{E} \nonumber
\end{eqnarray}
\end{widetext}
gives the distribution of the energy along the chain.

\section{Dynamical properties of solitons}
At the beginning let us consider stationary soliton solutions of
the problem (\ref{f13}). In the dimensionless functional
$\bar{L}$, coefficients $c_\alpha=c_\beta=0$ when $v=0$. So, only
one dimensionless parameter $g$ (\ref{f12}) which characterizes
cooperativity of rotational motions remains in functional
(\ref{f11}). The existence of soliton solution and its form depend
on the value of the parameter.

\subsection{Stationary solution}
The results of numerical investigations of the problem (\ref{f13}) show
that in the homogeneous chains stationary topological soliton solutions
exist when the parameter of cooperativity $g$ is larger than the threshold
value: $g\ge g_0>0$. The absence of the soliton topological stability when
$g<g_0$ can be explained in the following way. Any topological defect can
be eliminated by turning the DNA bases. The turning of one base
about 360 degrees transfers the system to the initial state, that is
$\phi_{i,n}\equiv\phi_{i,n}\pm 2\pi$. And this is why
the narrow solitons with the size equal to one link of the chain
are equivalent to the ground state and this is why they are unstable.
So, only relatively wide solitons with the general turning consisting
of several small changes of rotational angles that is when
$|\phi_{i,n+1}-\phi_{i,n}|\ll 2\pi$, are stable.
Dependence of the threshold value $g_0$ on the
soliton topological charge {\bf q} for homogeneous AT and GC
chains are given in the Table \ref{tab3}.
%-------------------------------------------------------------------------
\begin{table}[b]
\caption{\label{tab3}
Dependence of the threshold value of the parameter of cooperativity $g_0$
on the value of the soliton topological charge {\bf q}$=(q_1,q_2)$
for homogeneous $\alpha\beta=$AT (GC) chain}
\begin{ruledtabular}
\begin{tabular}{lccr}
$(q_1,q_2)$ &  (1,0)  & (0,1) & (1,1) \\
\hline
AT &  8.3   &   5.7  &   8.3  \\
GC &  12.0  &   8.2  &   12.0 \\
\end{tabular}
\end{ruledtabular}
\end{table}
%-------------------------------------------------------------------------
%-------------------------------------------------------------------------
\begin{table}[b]
\caption{\label{tab4} Dependence of the energy $E$ and the
diameter $D$ of stationary topological soliton on its topological
charge {\bf q} with two values of the transverse rigidity $K$}
\begin{ruledtabular}
\begin{tabular}{lccccr}
~&~&\multicolumn{2}{c}{$\alpha\beta=$AT}&\multicolumn{2}{c}{$\alpha\beta=$GC}\\
\hline
$K$ (N/m) &{\bf q} & $E$ (kJ/mol) & $D$ & $E$ (kJ/mol) & $D$  \\
\hline
     ~&(1,0) & 2776.52 & 16.59 &  3405.48 & 13.69 \\
0.234&(0,1) & 2237.36 & 19.58 &  2733.60 & 16.18\\
     ~&(1,1) & 4971.72 & 42.85 &  6087.54 & 35.13\\
\hline
     ~&(1,0) & 5329.76 &  9.03 &  6394.96 &  7.64 \\
0.8714&(0,1) & 4302.09 & 10.59 &  5146.38 &  8.98\\
     ~&(1,1) & 9551.14 & 22.60 & 11444.96 & 18.93\\
\end{tabular}
\end{ruledtabular}
\end{table}
%-------------------------------------------------------------------------

\begin{figure*}[htb]
\centering\epsfig{file=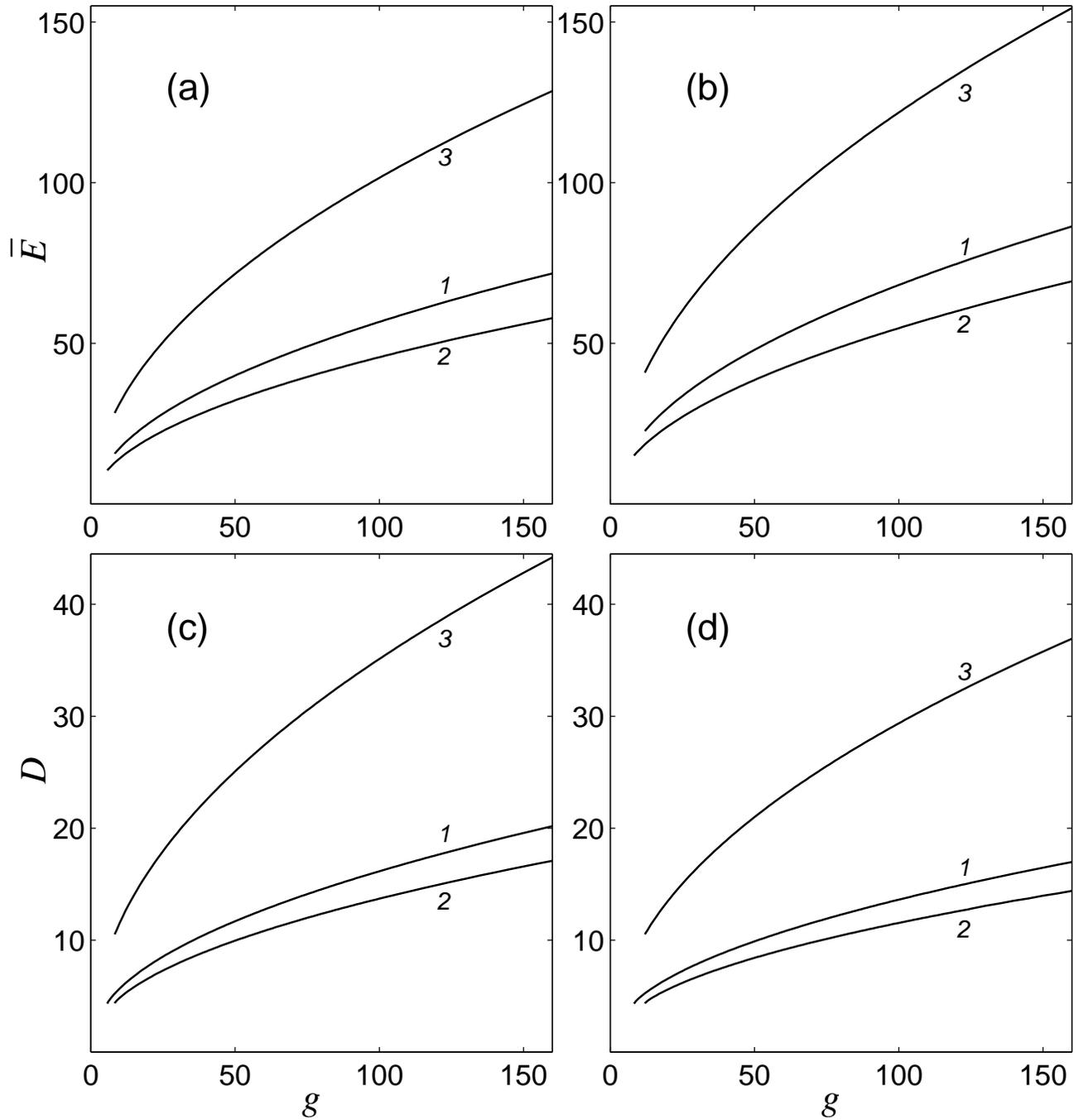,width=0.99\linewidth}
\caption{\label{fig3}\protect
        Dependence of the dimensionless soliton energy $\bar{E}$ and the width $D$
        on the velue of the parameter of cooperativity $g$ when {\bf q}=(1,0), (0,1),
        (1,1) (curves 1, 2, 3) in homogeneous AT (a),(c) and GC chains (b),(d).}
 \end{figure*}

\begin{figure}[htb]
\centering\epsfig{file=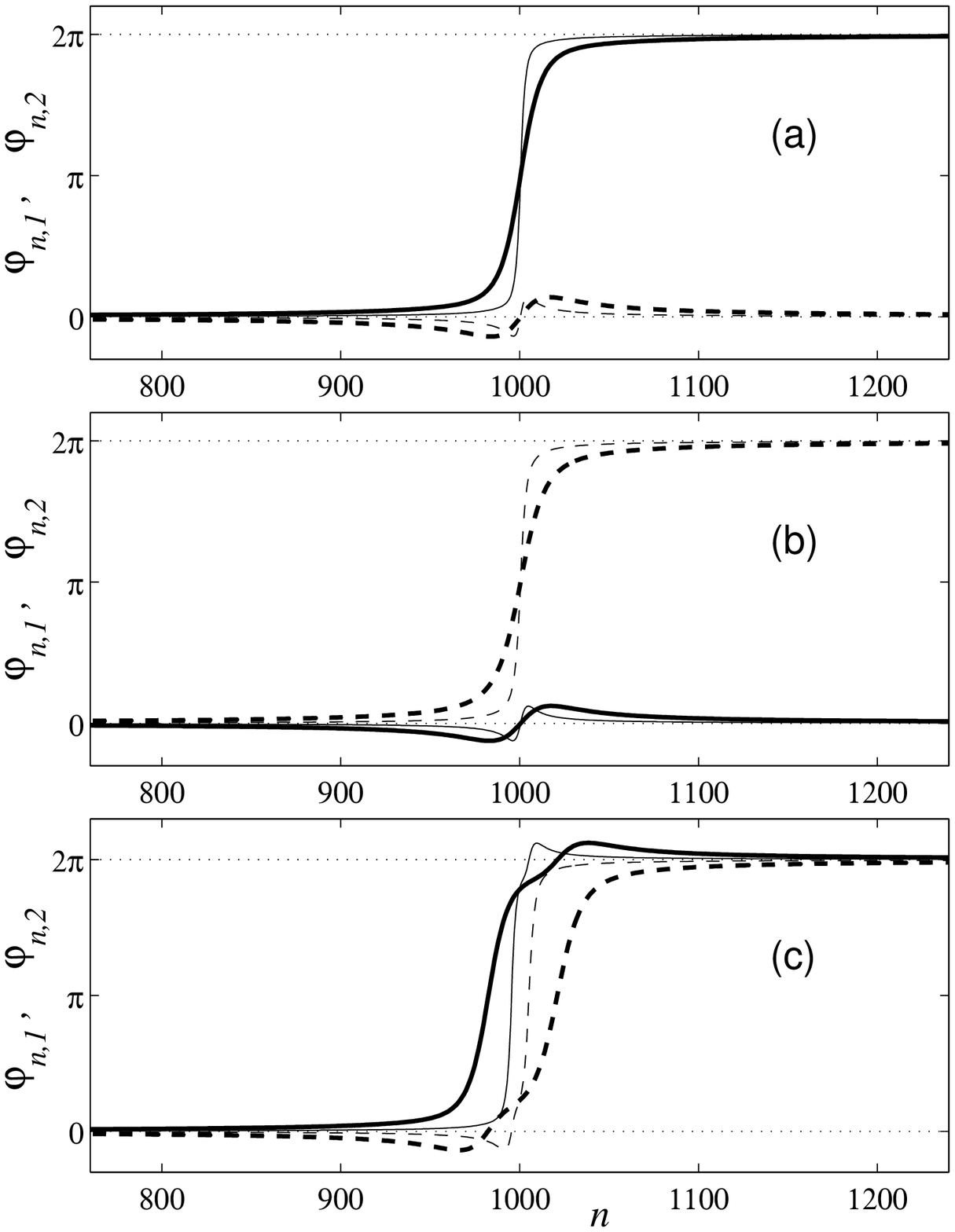,width=0.99\linewidth}
\caption{\label{fig4}\protect
         The view of stationary soliton with the topological charge {\bf q}=(1,0)
        (a); {\bf q}=(0,1) (b);  {\bf q}=(1,1)$_l$ (c).
        Continuous lines correspond to displacements by the first component
        $\varphi_{n,1}$; dotted lines -- to displacements by the second
        component $\varphi_{n,2}$; thin lines correspond to chain with
        $g=10$; flat lines -- to chain with $g=150$.
        }
 \end{figure}

From Fig. \ref{fig3} it becomes clear that the soliton energy
$\bar{E}$ and its width $D$ monotonously increase when the
parameter of cooperativity $g$ increases. For soliton stability it
is necessary that its width $D>4.33$. The view of stationary
solitons with the parameters of cooperativity $g=10$ and $g=150$
is presented in Fig. \ref{fig4}. In the case of the soliton with
topological charge {\bf q}=(1,0), the first component has the form
of a smooth step (when $n$ monotonously changes, the base of the
first of two DNA chains makes a complete turn) accompanied by
smooth small amplitude deformation in the second component (Fig.
\ref{fig4}a). In the case of the soliton with {\bf q}=(0,1), only
the second component has the form of a step (Fig. \ref{fig4}b). In
the case of soliton with {\bf q}=(1,1), each of the components has
the form of a step (Fig. \ref{fig4}c), steps being displaced
relatively one another. Later on we shall show that this soliton
is the bound state of two topological solitons with the charges
{\bf q}$_1$=(1,0) and {\bf q}$_2$=(0,1). There exist two
equivalent states of the soliton: the left state {\bf
q}=(1,1)$_l$, when the soliton with the charge {\bf q}$_1$ is on
left of the soliton with the charge {\bf q}$_2$ (Fig.
\ref{fig4}c), and the right state {\bf q}=(1,1)$_r$, when the
soliton with the charge {\bf q}$_1$ is on right of the soliton
with the charge {\bf q}$_2$.
%-------------------------------------------------------------------------
%-------------------------------------------------------------------------

When $\epsilon=6000$~kJ/mol and $K=0.234$~N/m the parameter of
cooperativity $g=150.24\gg g_0$, and when $K=0.8714$~N/m the
parameter $g=40.34>g_0$ (see Table \ref{tab3}) for all types of
topological solitons. So, for these values of the rigidity
parameter $K$ stable solitons with different topological charges
exist. For maximum value of the rigidity parameter $K=4.744$~N/m
the parameter of cooperativity $g=7.41<g_0$, and this means that
stable topological solitons are absent. Thus the problem of the
existence of topological solitons (open states) in the DNA reduces
to the problem of receiving exact estimation of the parameters
$\epsilon$ ш $K$. But this is, however, a rather difficult
problem. We think that the values $\epsilon=6000$~kJ/mol,
$K=0.8714$~N/m, when all three types of solitons exist, are the
most grounded. And we shall use these values for further
calculations.
%-------------------------------------------------------------------------
%-------------------------------------------------------------------------

Dependence of the energy $E$ and the diameter $D$ of stationary
topological soliton on its topological charge {\bf q} in the chain
with $\epsilon=6000$ kJ/mol and the transverse rigidity determined
by formulas (\ref{f4}) and (\ref{f8}), are given in the Table
\ref{tab4}. From the data of the table it follows that the energy
of the interaction of the solitons with the charges (1,0) and
(0,1) is equal to $\Delta E=E(1,0)+E(0,1)-E(1,1)=90.71$ kJ/mol for
the chain with the rigidity of the transverse interaction
$K=0.8714$ N/m, and the energy is equal to $\Delta E=42.16$ kJ/mol
for the chain with $K=0.234$ N/m.

\subsection{Nonstationary solutions}
Numerical investigation of the problem (\ref{f13}) shows, that in
the homogeneous chain, topological soliton has the interval of the
velocities $0\le s\le s_1<1$, where $s=v/v_0$ is the dimensionless
velocity and $v_0$ is the velocity of sound. Dependence of maximum
velocity of the soliton $s_1$ on its charge {\bf q}, on type of
the base in the chain $\alpha\beta$ and on the rigidity of the
transverse interaction $K$ is presented in the Table \ref{tab5}.
%-------------------------------------------------------------------------
\begin{table}[t]
\caption{\label{tab5}
Dependence of the maximum value of the soliton velocity $s_1$
on its topological charge {\bf q}, the soliton moving in the
homogeneous $\alpha\beta$ chain with the transverse
rigidity $K$.}
\begin{ruledtabular}
\begin{tabular}{lccr}
$K$ (N/m) &{\bf q}&  AT &  GC \\
\hline
      & (1,0) &   0.77  &  0.70 \\
0.234 & (0,1) &   0.88  &  0.84 \\
      & (1,1) &   0.77  &  0.70 \\
\hline
      & (1,0) &  0.64   &  0.55 \\
0.8714 & (0,1) &  0.86   &  0.84 \\
      & (1,1) &  0.65   &  0.56 \\
\end{tabular}
\end{ruledtabular}
\end{table}
%-------------------------------------------------------------------------

Dependence of the soliton energy $E$ and the diameter $D$ on the
dimensionless velocity $s$ is presented in Fig. \ref{fig5}.
With the increasing of the soliton velocity its energy monotonically
increases, and the diameter monotonically decreases.
Using the dependence $E(s)$ we can find the mass of rest
of the topological soliton
$$
M=\lim_{s\rightarrow 0}\frac{2(E(s)-E(0))}{s^2v_0^2}~.
$$
Dependence of the mass of rest $M$ of the soliton on its charge {\bf q},
on the type of the bases of the chain $\alpha\beta$ and on the rigidity
of the transverse interaction $K$ is presented in the Table {\ref{tab6}.
%-------------------------------------------------------------------------
\begin{figure}
\centering\epsfig{file=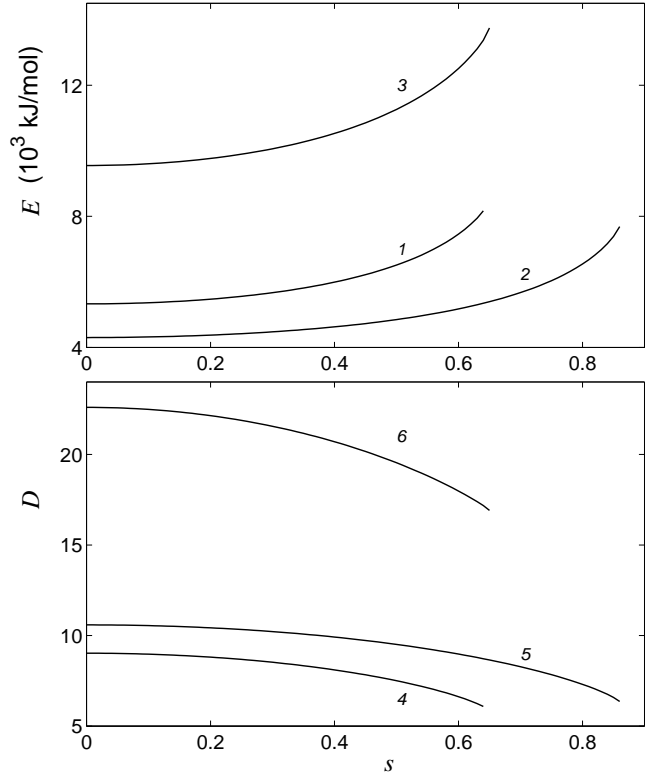,width=0.99\linewidth}
\caption{\label{fig5}\protect
        Dependence of the energy $E$ and the diameter $D$ of the
        soliton which moves along homogeneous AT chain and has the charge {\bf q}$=(1,0)$
        (curves 1 and 4), {\bf q}$=(0,1)$ (curves 2 and 5) and {\bf q}$=(1,1)$
        (curves 3 and 6), on the dimensionless velocity $s$
        ($\epsilon=6000$ kJ/mol, $K=0.8714$ N/m).
        }
 \end{figure}
%-------------------------------------------------------------------------
%-------------------------------------------------------------------------
\begin{table}[t]
\caption{\label{tab6}
The dependence of the soliton mass of rest $M$ (the values are
given in proton mass units
$m_p$) on its topological charge {\bf q}, the soliton being
in the homogeneous $\alpha\beta$ chain with the transverse
rigidity $K$.}
\begin{ruledtabular}
\begin{tabular}{lccr}
$K$ (N/m) &{\bf q}&  AT &  GC \\
\hline
      & (1,0) &   7640  &  9663 \\
0.234 & (0,1) &   4052  &  4064 \\
      & (1,1) &   11581  &  13590 \\
\hline
      & (1,0) &  14978   &  18730 \\
0.8714 & (0,1) &  7899   &   7804 \\
      & (1,1) &  22660   &  26265 \\
\end{tabular}
\end{ruledtabular}
\end{table}
%-------------------------------------------------------------------------

Numerical investigation shows that all topological solitons
at all permitted velocities are stable.
They move along the chain with constant velocity, their
form and energy being conserved.
Thus, the specificity of the chain of the DNA molecule
leads to a principal effect consisting in the possibility
of preferable localization of soliton excitations on one chain.
Moreover, it appears that soliton excitation with the charge (1,1)
is a bound state of two excitations localized in separate chains.

\section{Interaction of topological solitons}
DNA is a rather long molecule, and several open states can be activated
in it simultaneously. Therefore it is
interesting to consider the problem of interaction of solitons imitating the open states.

Numerical approach
(\ref{f13}) permits us to investigate the problem
and to obtain the dependence of the energy
of a pair of solitons with the charges {\bf q}$_1$,
{\bf q}$_2$ on the distance between their centers $n_1$, $n_2$.
For the purpose, it is necessary to take
the boundary conditions and the initial
point, which correspond to a pair of topological solitons with
the centers moving away at a distance $R$. When minimizing the energy
of the system $E=-L$, it is necessary also to fix the turns of the bases,
which correspond to the centers of solitons. Then the energy
of the obtained state $E(R)$ corresponds to the energy of a pair
of topological solitons, the solitons being at a distance $R=n_2-n_1$
from each other. By changing positions of the soliton centers,
we can obtain the potential of interaction
$$
U_{{\bf q}_1,{\bf q}_2}(R)=E(R)-E({\bf q_1})-E({\bf q_2})~,
$$
where $E({\bf q_1})$ and $E({\bf q_2})$ is the energy of isolated
solitons.

The potential of interaction of solitons of different types and
the potential of topological solitons with the charges of the same
sign are presented in Fig. \ref{fig6}. The potential of two solitons
of different types with {\bf q}$_1=(1,0)$ and
{\bf q}$_2=(0,1)$ has the form of symmetrical double well
potential (Fig. \ref{fig6}, curve 1). Maximum of the potential
is reached when $R=0$, that is when the centers
of the solitons are placed at the neighboring chains and when the solitons are
opposite to each other. From energetic point of view this configuration
of the solitons of different chains is the most
disadvantageous.
Minimum of the energy is reached when
$R=\pm20a$. Thus two solitons of this type can form
two energetically equivalent coupled states. One
of the states (left minimum of the potential of interaction)
corresponds to the left isomer of the topological soliton
with the charge {\bf q}$=(1,1)_l$, and the other state
(the right minimum of the potential of interaction) corresponds
to the right isomer of the soliton ${\bf q}=(1,1)_r$.
%-------------------------------------------------------------------------
\begin{figure}
\centering\epsfig{file=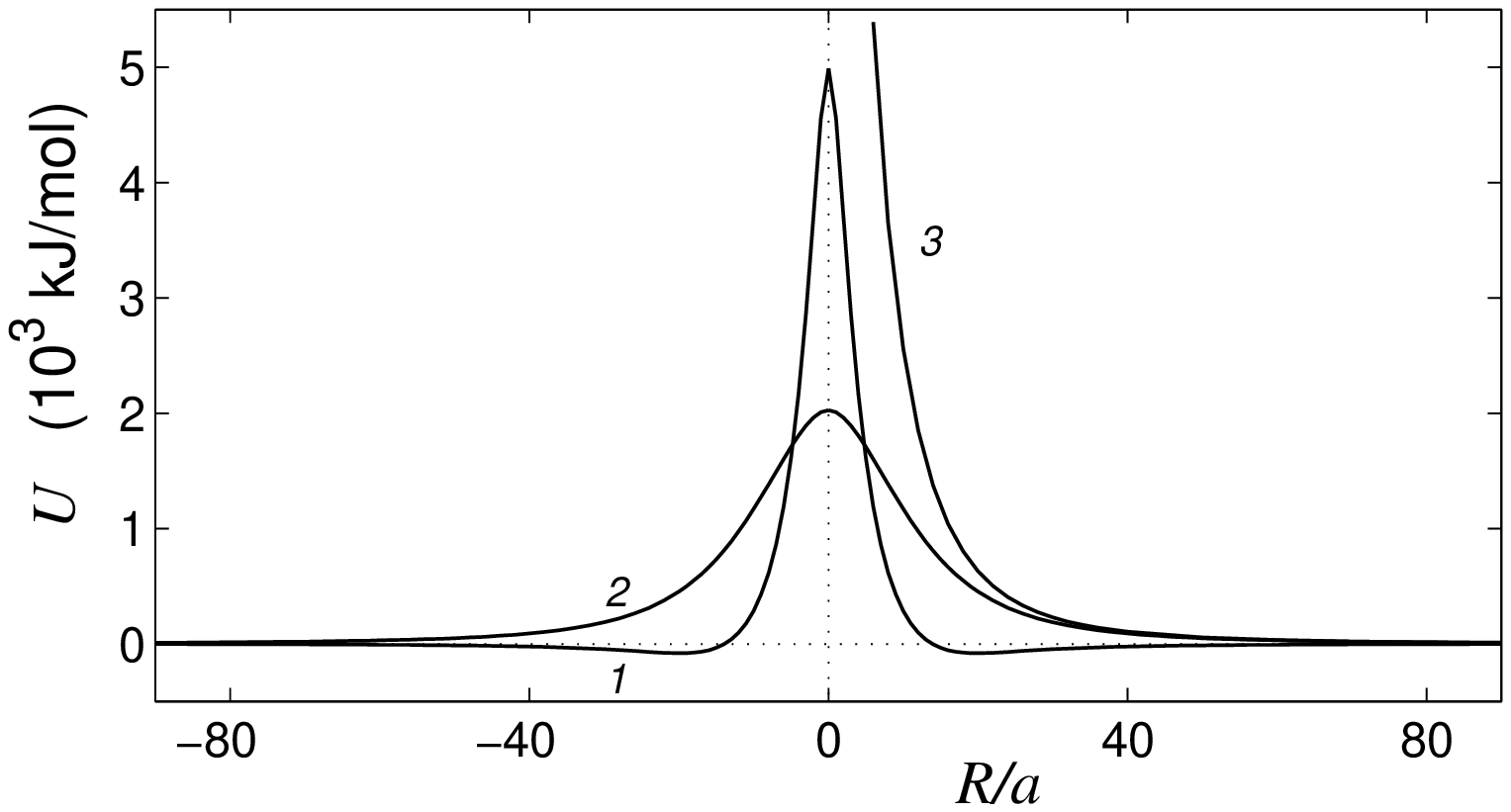,width=0.99\linewidth}
\caption{\label{fig6}\protect
        Potential of interaction of solitons $U_{{\bf q}_1,{\bf q}_2}(R)$
        with {\bf q}$_1=(1,0)$, {\bf q}$_2=(0,1)$ (curve 1);
        {\bf q}$_1=(1,0)$, {\bf q}$_2=(0,-1)$ (curve 2);
        {\bf q}$_1={\bf q}_2=(1,0)$ and {\bf q}$_1={\bf q}_2=(0,1)$
        (curve 3).
        }
 \end{figure}
%-------------------------------------------------------------------------

If solitons have different signs of charges {\bf q}$_1=(1,0)$
and {\bf q}$_2=(0,-1)$, the potential of interaction has a
bell-like form with one maximum
at $R=0$ (Fig. \ref{fig6}, curve 2). From the potential it follows
that, the solitons which belong to different chains,
should repulse from each other. Solitons with the same sign of charge
({\bf q}$_1={\bf q}_2=(1,0)$, (0,1)) also repulse from each other.

When distance between
the solitons decreases, the energy monotonically increases
and goes to infinity when $R\rightarrow 0$ (Fig. \ref{fig6}, curve 3).

The potential of interaction $U_{{\bf q}_1,{\bf q}_2}(R)$ permits
to predict the result of repulsion of solitons with the charges
${\bf q}_1$ and ${\bf q}_2$. Let us model the repulsion of the solitons.
For the purpose, let us consider a double chain consisting of $N=4000$
base pairs. At the ends of each of the polynucleotide chains let us introduce
viscous friction which provides with absorption of phonons.
The system of equations of motion
(\ref{f5}), $n=1,2,...,N$, was integrated numerically
with the initial condition
which corresponds to two topological solitons with the centers placed
in the points $n_1=N/4$ and $n_2=3N/4$  and with the velocities
$s_1=-s_2=s>0$.

The results show that collision of solitons having equal signs
{\bf q}$_1={\bf q}_2=(\pm 1,0),~(0,\pm 1)$
leads to their reflection at one another.
When the velocities $s$ are small,
the reflection is practically elastic, and when the velocities
$s$ are large, collision is accompanied by slight emission of phonons.
Collision of solitons with {\bf q}$_1=(\pm 1,0)$, {\bf q}$_2=(0,\mp 1)$
and $s=0.5$ leads to their reflection, accompanied by slight emission
of phonons. This behavior is in a good agreement with the form
of corresponding potential of interaction (Fig. \ref{fig6}, curve 2).
To pass through one another, solitons need to overcome
energy barrier $U_{{\bf q}_1,{\bf q}_2}(0)=2025$ kJ/mol.
Thus, their kinetic energy should be equal to
$E_k(s)=E_{{\bf q}_1}(s)+E_{{\bf q}_2}(s) -E_{{\bf q}_1}(0)
-E_{{\bf q}_2}(0)>U_{{\bf q}_1,{\bf q}_2}(0)$. This condition is
fulfilled only in the vicinity of the most possible values of the velocity
(see. Fig. \ref{fig5}).
So, when $s=0.5$, the kinetic energy $E_k=1732$ kJ/mol is lower than
the height of the energy barrier (reflection takes place), and when $s=0.6$,
the energy $E_k=3000.9$ kJ/mol is higher than the barrier
(solitons pass through one another).

Solitons with the charges {\bf q}$_1=(\pm 1,0)$, {\bf q}$_2=(0,\pm
1)$ attract one another at a distance $R>20a$, and when the
distance is shorter they repulse one another. Here the energy
barrier $U_{{\bf q}_1,{\bf q}_2}(0)=4989$ kJ/mol does not permit
solitons to pass through one another. Solitons always reflect.
Formation of the bound state does not occur even when the value of
the velocity is small. It is explained by small value of the bond
energy $\Delta E=91$ kJ/mol.

Potential of interaction of different topological solitons
with the charges of opposite signs
$({\bf q}_1=-{\bf q}_2)$ is presented in Fig. \ref{fig7}.
Potential of interaction of one component solitons
$({\bf q}_1=(1,0),~(0,1))$
monotonically decreases with the decreasing of the distance between
the solitons. When $R\rightarrow 0$, the potential
$U_{{\bf q}_1,{\bf q}_2}(R)\rightarrow -[E({\bf q}_1)+E({\bf q}_2)]$.
At a distance $R=0$ solitons completely recombine.
%-------------------------------------------------------------------------
\begin{figure}
\centering\epsfig{file=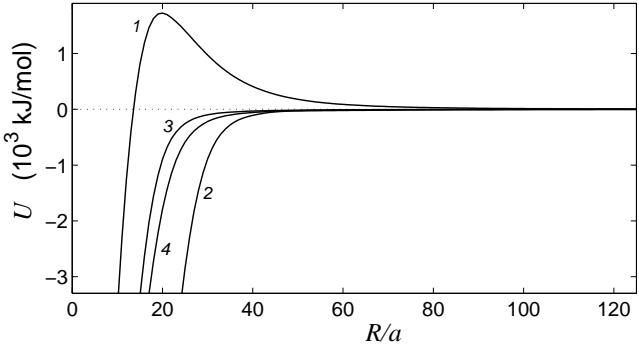,width=0.99\linewidth}
\caption{\label{fig7}\protect
        Potential of interaction of solitons $U_{{\bf q}_1,{\bf q}_2}(R)$
        with {\bf q}$_1=(1,1)_l$, {\bf q}$_2=(-1,-1)_l$ (curve 1);
        {\bf q}$_1=(1,1)_l$, {\bf q}$_2=(-1,-1)_r$ (curve 2);
        {\bf q}$_1=(1,0)$, {\bf q}$_2=(-1,0)$ (curve 3);
        {\bf q}$_1=(0,1)$, {\bf q}$_2=(0,-1)$ (curve 4).
        }
 \end{figure}
%-------------------------------------------------------------------------

Potential of interaction of two component
solitons with the charges of different signs has a similar form.
If the solitons have different polarity, that is, if the first soliton
is a left isomer $({\bf q}_1=(1,1)_l)$ and the second soliton is
a right isomer $({\bf q}_2=(-1,-1)_r)$. Solitons of that type
attract one another. Their collisions always leads to
recombination of the solitons.

If solitons have the same polarity, they repulse when $R>20a$, and
attract when the distances are shorter. Recombination of the solitons
requires overcoming the energy barrier 1730 kJ/mol.
Solitons overcome the barrier only when $s>0.38$,
and the value of their kinetic energy is more than the height
of the barrier. When the value of the velocity is smaller,
solitons reflect, and when the value is larger they recombine
(Fig. \ref{fig8}). During recombination the energy of solitons
is spent for intensive emission of phonons. Breather-like excitations
can be also formed.
%-------------------------------------------------------------------------
\begin{figure}
\centering\epsfig{file=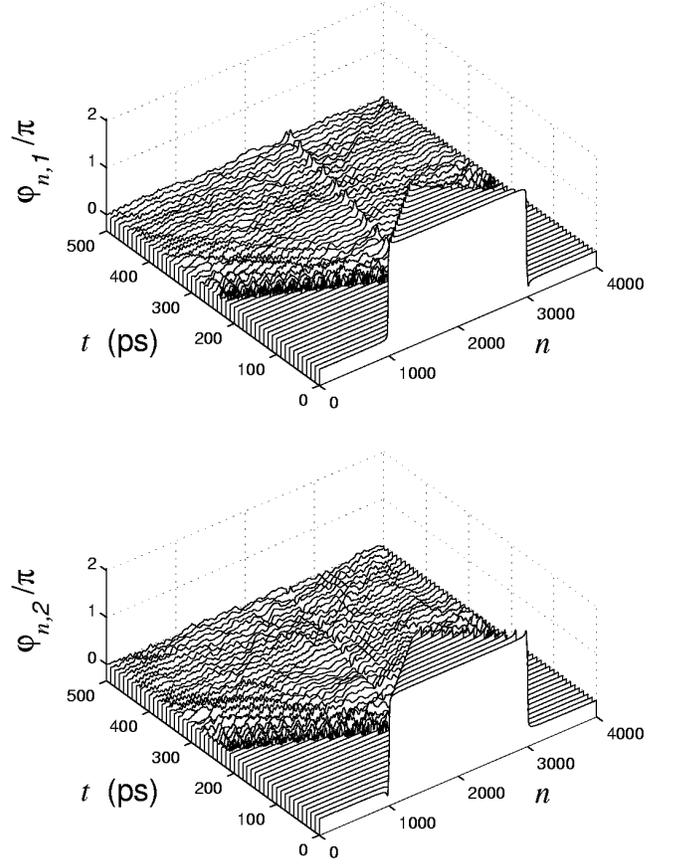,width=0.99\linewidth}
\caption{\label{fig8}\protect
        Recombination of solitons with the same signs of charges and the same polarities
        (${\bf q}_1=(1,1)_l$, ${\bf q}_2=(-1,-1)_l$, $s=0.5$).
        }
 \end{figure}
%-------------------------------------------------------------------------

Collision of one component soliton with two component soliton can lead (depending on the
relationship of the signs of charges and on the polarity of the
two component soliton) to their partial
recombination or to inelastic reflection
accompanied by disintegration of the two component soliton.
So, at the velocity
$s=0.5$ collision of the soliton having the charge
${\bf q}_1=(1,0)$, with  the two component
soliton having the charge ${\bf q}_2=(1,1)_l$, leads
to inelastic reflection of the first
soliton, and at the same time the second soliton disintegrates
into two one component solitons
with the charges (1,0) and (0,1) (Fig. \ref{fig9}).
%-------------------------------------------------------------------------
\begin{figure}
\centering\epsfig{file=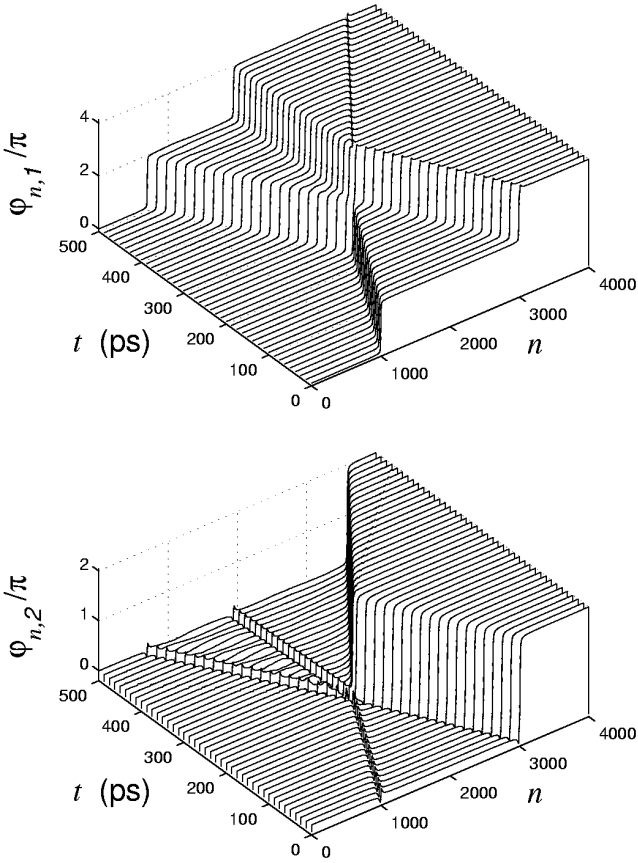,width=0.99\linewidth}
\caption{\label{fig9}\protect
         Disintegration of two component soliton (${\bf q}_2=(1,1)_l)$ when
        it comes into collision with one component soliton (${\bf q}_1=(1,0))$.
        The velocity of the movement is $s=0.5$.
        }
 \end{figure}
%-------------------------------------------------------------------------

\section{Effect of the chain inhomogeneities on the dynamics of
         topological solitons}
Till now our investigation was limited by consideration of
homogeneous model of DNA. The real DNA is, however, a
substantially inhomogeneous system, therefore it is of special
interest to consider the effect of the chain inhomogeneity on the
dynamics of topological solitons. In the inhomogeneous chain, the
energy of stationary soliton $E$ will depend on the position of
the center of the soliton $\bar{n}$. To move, soliton requires to
overcome the energetic potential barrier $E(\bar{n})$. To find the
energy of the soliton with the center at the point $n=\bar{n}$ we
need to solve numerically the problem of minimizing
\begin{eqnarray}
E\rightarrow \min_{\varphi_{2,i},...,\varphi_{N-1,i},~i=1,2}:
 \label{f16}\\
\varphi_{1,1}=\varphi_{-\infty,1},~~
\varphi_{1,2}=\varphi_{-\infty,2},\label{f17} \\
\varphi_{N,1}=\varphi_{\infty,1},~~
\varphi_{N,2}=\varphi_{\infty,2}, \label{f18}
\end{eqnarray}
where the energy
\begin{eqnarray}
E=\sum_n [
\epsilon(\sin^2\frac{\varphi_{n+1,1}-\varphi_{n,1}}{2}
+\sin^2\frac{\varphi_{n+1,2}-\varphi_{n,2}}{2})\nonumber \\
+V_{\alpha\beta_n}(\varphi_{n,1},\varphi_{n,2})], \nonumber
\end{eqnarray}
$\alpha\beta_n$ -- the sequence of the base pairs along the chain.
The boundary conditions (\ref{f17}), (\ref{f18}) are the same as
those in the problem (\ref{f13}). To fix the soliton center
position we need to solve the problem of minimizing (\ref{f16})
with respect to variables $\varphi_{n,i}$ where $2\le n\le N-1$,
$n\neq\bar{n}$, $i=1,2$. For soliton with charge $q_1\neq 0$ it is
necessary to fix the value $\varphi_{\bar{n},1}=\pi q_1$, and for
soliton with $q_2\neq 0$ -- the value $\varphi_{\bar{n},2}=\pi
q_2$.

The problem on conditional minimum (\ref{f16}) has been solved by
the numerical method of conjugate gradient. We took $N=2000$.
Possible view of energetic profile of the soliton moving in the
inhomogeneous chain is presented in Fig. \ref{fig10}
%-------------------------------------------------------------------------
\begin{figure}
\centering\epsfig{file=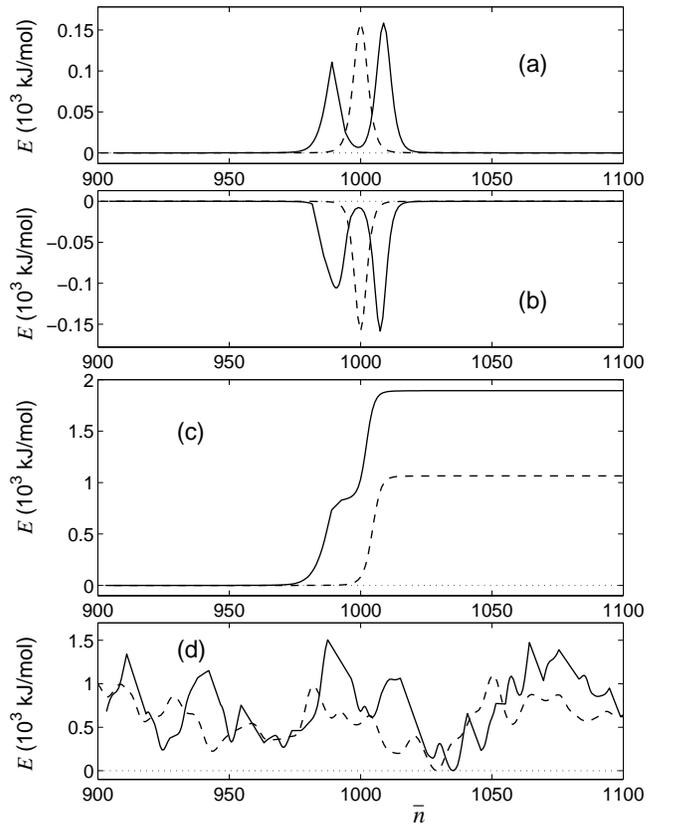,width=0.99\linewidth}
\caption{\label{fig10}\protect
        The view of energetic relief for soliton in the inhomogeneous chain:
        in homogeneous AT chain with one GC base pair (a); in homogeneous
        GC chain with one AT base pair (b); in the chain the first part of
        which consists of only AT base pairs, and the second - of only
        GC base pairs (c); in the chain with random sequence of base pairs
        (d). Dotted line shows the relief of the soliton with topological
        charge {\bf q}$=(1,0)$, firm line -- the relief for soliton with
        charge {\bf q}$=(1,1)$.
        }
 \end{figure}
%-------------------------------------------------------------------------

At the beginning, let us estimate the effect of point
inhomogeneities. From Fig. \ref{fig10}a it is obvious that one
point defect in the homogeneous AT chain leads to appearance of
localized potential barrier with height equal to $E_d=150$ kJ/mol.
To overcome the barrier, the soliton kinetic energy should satisfy
the condition $E_k(s)=E(s)-E(0)>E_d$. Soliton can to overcome the
barrier only when its velocity $s>s_d$, where the threshold value
of the velocity $s_d$ is taken from equation $E(s_d)-E(0)=E_d$.
From the data of Fig. \ref{fig5} it is easy to find, that for
soliton with {\bf q}=(1,0) the velocity $s_d=0.21$, for soliton
with {\bf q}=(0,1) the velocity $s_d=0.28$, and when {\bf q}=(1,1)
the velocity $s_d=0.17$.

Let us model numerically the interaction of soliton with local
defect of the homogeneous AT chain. For the purpose, let us
consider homogeneous AT chain consisting of $N=4000$ bases with
one base GC in the middle of the chain in the point $n=N/2$.
Suggest that at the initial time a topological soliton is in the
point $n=N/4$ and consider its movement through the chain
inhomogeneity. The results of numerical modeling of the soliton
dynamics show, that independently on the value of topological
charge {\bf q} soliton with velocity $s=0.05$ reflect from this
point defect, but for $s=0.5$, moves through the point defect with
negligibly small energy loss.

Point defect in the homogeneous GC chain leads to formation of
localized potential well with depth 150 kJ/mol (see Fig.
\ref{fig10}b). Almost at all values of the velocity soliton easily
propagates along this chain without formation of bound state. Thus
we can conclude that soliton moving in the DNA chain with
sufficiently large velocity ($s>s_d$) is stable with respect to
point defects.

In the chain one part of which consists of only AT base pairs, and
the other -- of only GC base pairs energetic barrier takes the
form of smooth step (Fig. \ref{fig10}c). The height of the step is
equal to the difference between the values of soliton energy in
homogeneous GC and AT chains. From data of table \ref{tab4} it
follows, that the height of the step is equal to $\Delta E=1065$
kJ/mol for soliton with topological charge {\bf q}=(1,0) and
$\Delta E=844$ kJ/mol for soliton with {\bf q}=(0,1), and when
{\bf q}=(1,1) the energy $\Delta E$=1894 kJ/mol. Soliton moving
along homogeneous AT region of the chain, can enter into GC region
only if its kinetic energy $E_k(s)>\Delta E$. As seen from Fig.
\ref{fig5}, this condition is satisfied only if soliton velocity
$s>s_k$, where the threshold value of the velocity is determined
by the equation $E_k(s_k)=\Delta E$. For soliton with {\bf
q}=(1,0) the velocity $s_k=0.48$, for {\bf q}=(0,1) $s_k=0.59$,
and for {\bf q}=(1,1) the velocity $s_k=0.52$.
%-------------------------------------------------------------------------
\begin{figure}
\centering\epsfig{file=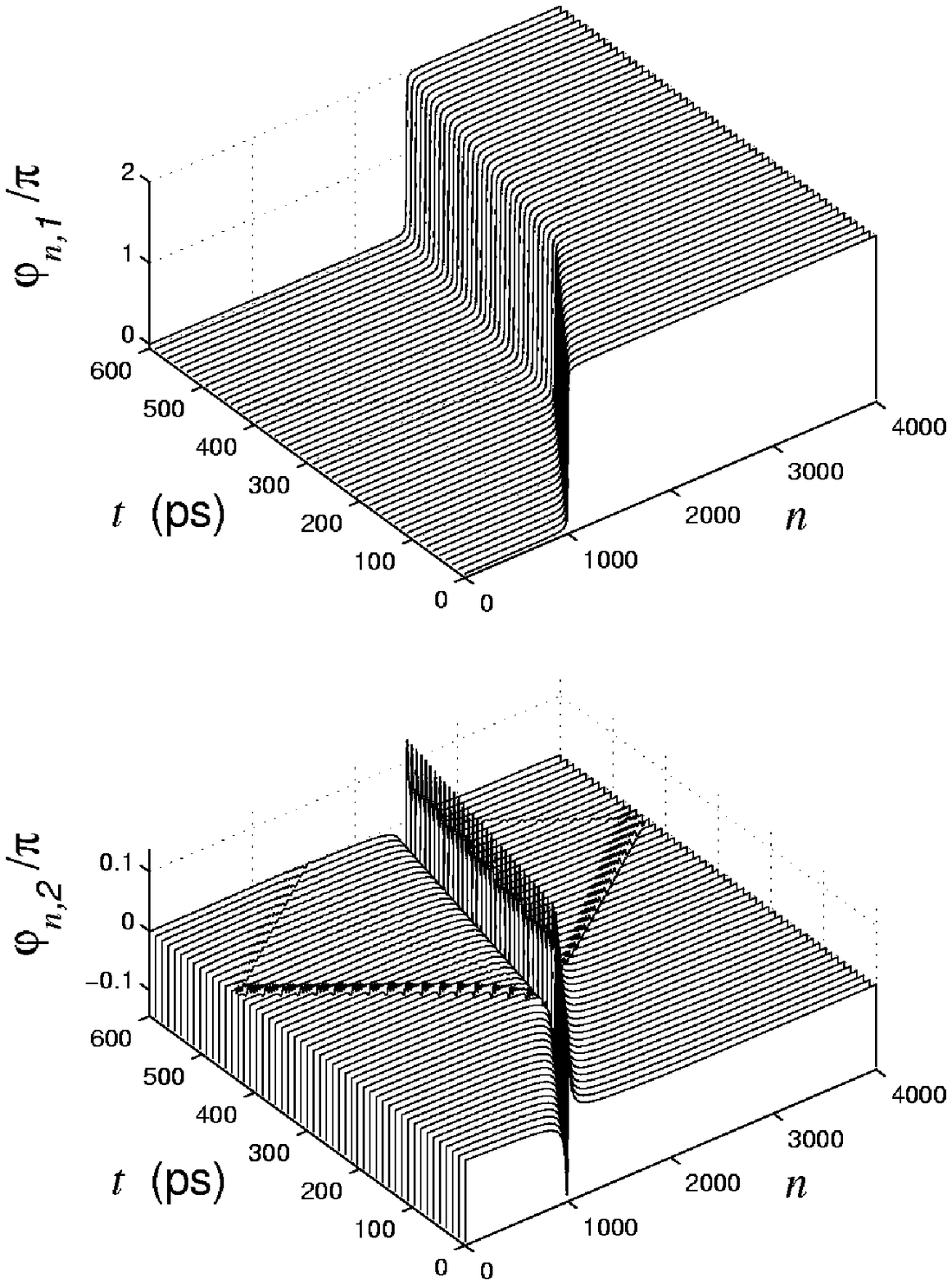,width=0.99\linewidth}
\caption{\label{fig11}\protect
        Movement of soliton with {\bf q}$=(1,0)$, $s=0.5$ through the
         boundary between homogeneous AT and GC regions.
        }
 \end{figure}
%-------------------------------------------------------------------------

Let us model numerically the soliton moving from homogeneous AT
region of the chain to homogeneous GC region. The results of the
modeling show that the soliton with velocity $s=0.05<s_k$ and with
any topological charge reflects elastically from the boundary
between the regions. At given velocity the soliton kinetic energy
is not large enough to overcome energetic barrier
($E_k(s)\ll\Delta E$). Soliton with $s=0.5$ ${\bf q}=(1,0)$ moves
through the boundary between homogeneous regions and its motion is
accompanied by emission of phonons. Inside the region consisting
of GC base pairs, soliton continues to move, but with a smaller
magnitude of the velocity (Fig. \ref{fig11}). For given value of
{\bf q} the threshold value of the velocity $s_k=0.48<0.5$. So,
the kinetic energy of soliton is large enough to overcome
energetic barrier. Because the main part of the kinetic energy is
spent to overcome the barrier, the velocity of the soliton
substantially decreases after overcoming the barrier. When {\bf
q}=(0,1) the threshold value of the velocity $s_k=0.59$ and
soliton reflects at $s=0.5$ from the boundary of the homogeneous
regions. The reflection is accompanied by phonon emission. For
soliton with the charge {\bf q}=(1,1) the velocity
$s=0.5<s_k=0.52$ is not enough to overcome energetic barrier.
Collision of soliton having topological charge $({\bf q}=(1,1)_l)$
with the boundary between the homogeneous regions leads to the
disintegration of the soliton. It disintegrates into two one
component solitons with the charges ${\bf q}_1=(1,0)$ and ${\bf
q}_2=(0,1)$. Soliton with the charge {\bf q}$_2$ continues to move
into GC region of the chain, and soliton with the charge {\bf
q}$_1$ reflects from the boundary.
%-------------------------------------------------------------------------
\begin{figure}
\centering\epsfig{file=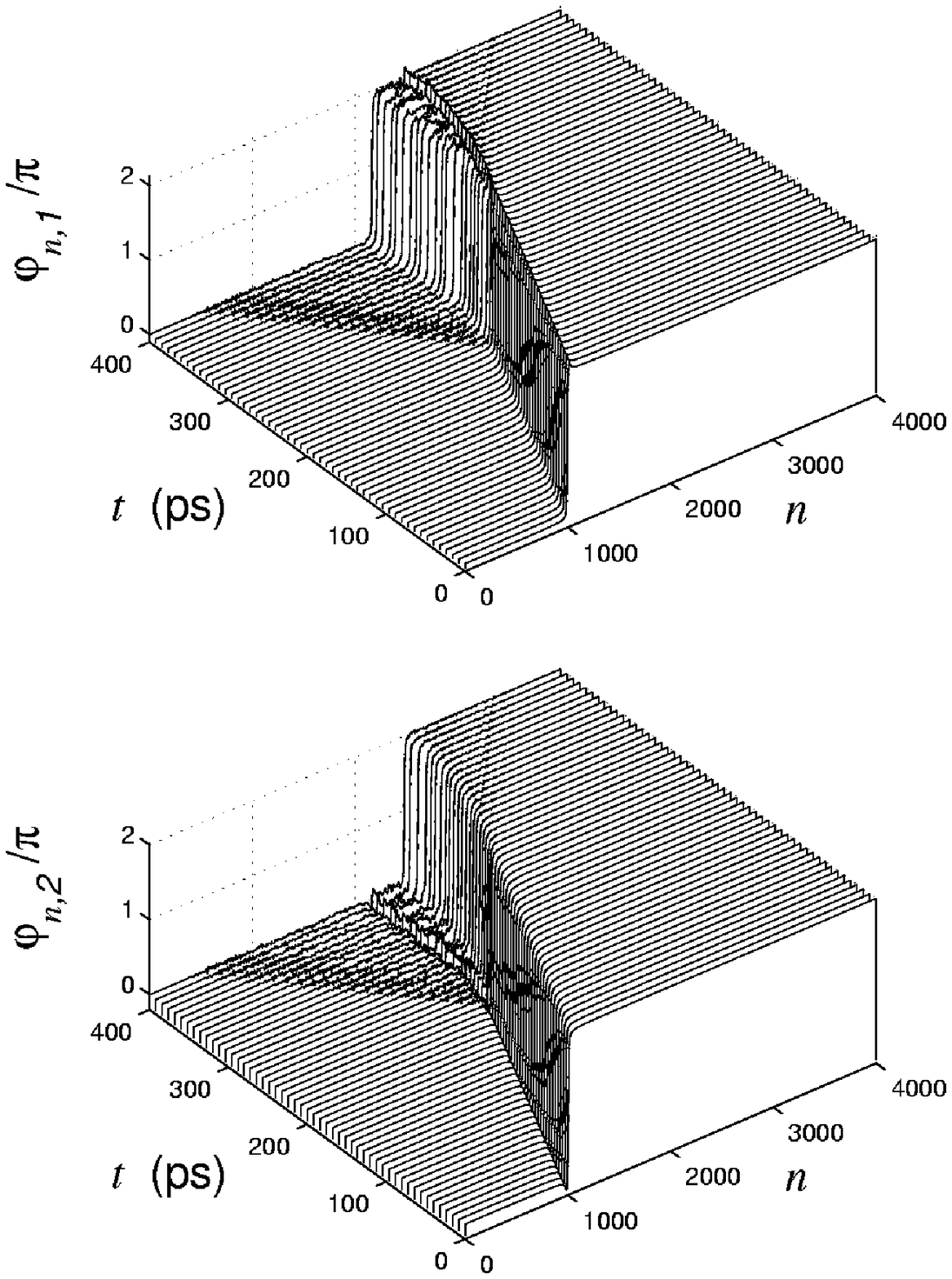,width=0.99\linewidth}
\caption{\label{fig12}\protect
         Entering two component soliton (charge
        {\bf q}$=(1,1)_l$, velocity $s=0.5$) the random inhomogeneous region of
        the chain, and further disintegration of the soliton.
        }
 \end{figure}
%-------------------------------------------------------------------------

Let us consider the propagation of soliton in the inhomogeneous chain
with random sequence of bases. In this case random energetic
relief $E(n)$ is formed. The amplitude of the relief for soliton
with {\bf q}=(1,0) reaches 1000 kJ/mol, and for soliton with {\bf
q}=(1,1) -- 1500 kJ/mol (Fig. \ref{fig10}d). It is obvious that
uniform propagation of soliton in the chain of that type is
impossible, because soliton loses part of energy for phonon emission
when crossing each homogeneity.

Let us consider the movement of soliton through inhomogeneous
region of the chain.
For the purpose, let us suggest that the second part
of the chain is formed by a random equal-possible sequence
of base pairs AT, TA, CG, GC. The results of numerical modeling
of the soliton dynamics show that soliton with small value
of the velocity $s=0.05$
and with any topological charge reflects from the boundary
of the inhomogeneous region. This points out, that penetration of the
soliton into the inhomogeneous region
requires the overcoming of some energy barrier. Soliton with larger
velocity $s=0.5$ and charge {\bf q}$=(1,0)$ overcomes this
barrier, enters the disordered region of the chain and stops there.
The movement in the disordered region is accompanied by intensive emission
of phonons, which leads to the stop of the soliton.
Soliton with {\bf q=}$(0,1)$ can not overcome the barrier even at this
value of the velocity. The soliton reflects from the boundary
of the inhomogeneous region.
The reflection is accompanied by emission of phonons.
Two component soliton with {\bf q}$=(1,1)_l$ enters inhomogeneous region,
and at the same time it disintegrates into two one component solitons
with the charges
{\bf q}$_1=(1,0)$ and {\bf q}$_2=(0,1)$. The solitons moves some
time in the inhomogeneous chain, then they stop (Fig. \ref{fig12}).
The path of the solitons can reach several hundred base pairs.

Analogous results have been obtained even in the case when inhomogeneous
region was formed by the base pairs AT and TA. Thus,
the sequence of nitrous bases of DNA molecule should substantially
influence the characteristics of the motion of topological soliton.
Note, that it has been pointed out firstly in the work \cite{p25}.

\section{Interaction of topological solitons with thermal
         oscillations of the chain}
Dynamics of a thermalized chain consisting of $N$ sites, is
described by the system of the Langevin equations
\begin{eqnarray}
I_{n,1}\ddot{\varphi}_{n,1} &=& -\frac{\partial H}{\partial\varphi_{n,1}}
+\xi_{n,1}-\Gamma I_{n,1}\dot{\varphi}_{n,1}~; \nonumber \\
I_{n,2}\ddot{\varphi}_{n,2} &=& -\frac{\partial H}{\partial\varphi_{n,2}}
+\xi_{n,2}-\Gamma I_{n,2}\dot{\varphi}_{n,2}~, \label{f19} \\
     n &=& 1,2,...N~, \nonumber
\end{eqnarray}
where the Hamiltonian of the system $H$ is given by Eq. (\ref{f1}),
$\xi_{n,i}$ are random normally distributed forces describing
the interaction of the $n$-th base of the $i$-th chain $(i=1,2)$ with
thermal bath,
$\Gamma=1/t_r$ is the coefficient of friction, $t_r$ being the relaxation time of
the rotation velocity of one base. The random forces $\xi_{n,i}$
have normal distribution and the correlation functions are
\begin{eqnarray}
\langle \xi_{n,i}(t_1)\xi_{m,j}(t_2)\rangle=2\Gamma k_B T
\delta_{nm}\delta_{ij}\delta(t_1-t_2)\sqrt{I_{n,i}I_{n,j}},
\nonumber \\
n,m= 1,2,...,N,~~i,j=1,2~~, \nonumber
\end{eqnarray}
where $k_B$ is  Boltzmann's constant and
$T$ is the temperature of thermal bath.

The system (\ref{f19}) was integrated numerically by the standard
fourth-order Runge-Kutta method with constant step of integration
$\Delta t$. The delta function was represented as $\delta (t)=0$
when $|t|>\Delta t/2$, and $\delta(t)=1/\Delta t$ when
$|t|\le\Delta t/2$, i.e., the step of numerical integration
corresponded to the correlation time of the random force. In order
to use the Langevin equation, it was necessary to suggest that
$\Delta t\ll t_r$. Therefore we chose $\Delta t=0.001$ ps and the
relaxation time $t_r\ge 1$ ps.
%-------------------------------------------------------------------------
\begin{figure}
\centering\epsfig{file=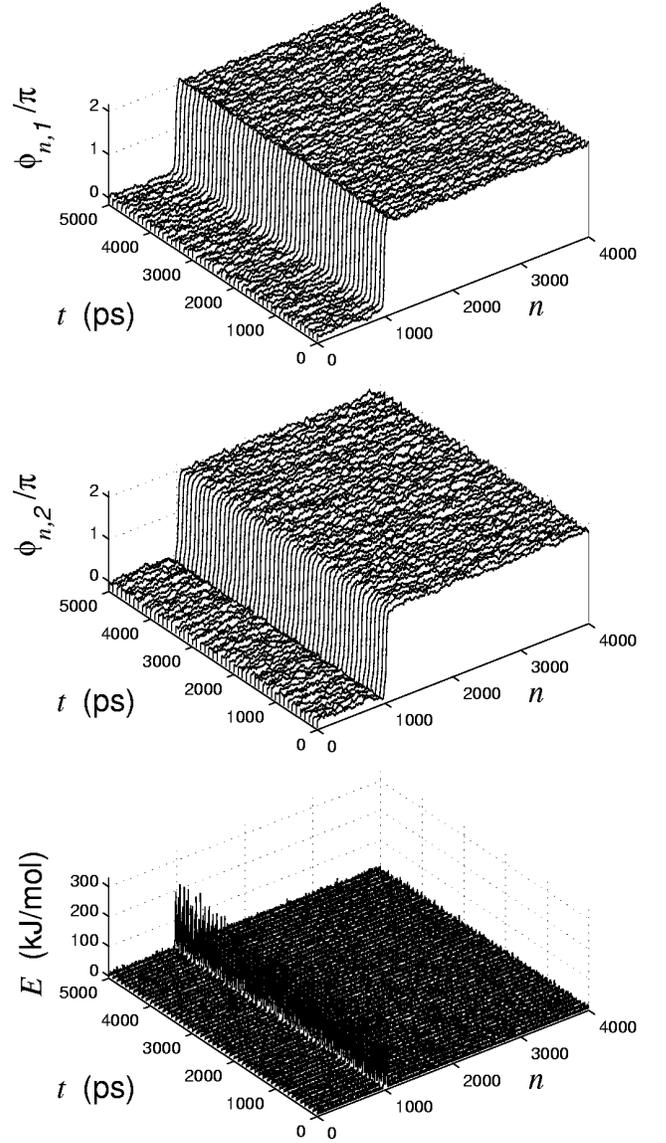,width=0.99\linewidth}
\caption{\label{fig13}\protect
        Stability of topological soliton ({\bf q}$=(1,1)_l$, $s=0.5$)
        in the thermalized homogeneous AT chain ($T=300$K, $t_r=1$ ps).
        Dependence of the distribution of angular displacements
        $\varphi_{n,1}$, $\varphi_{n,2}$ and energy $E_n$ along
        the chain on time $t$ is shown ($K=0.234$ N/m).
        }
 \end{figure}
%-------------------------------------------------------------------------

Let us check stability of topological soliton with respect to
thermal oscillations of the chain. For the purpose, let us
consider homogeneous periodical AT chain consisting of $N=4000$
base pairs at the temperature $T=300$K. Let us integrate system
(\ref{f19}) with the initial condition corresponding to
topological soliton $(s=0.5)$ with center placed in the point
$n=N/4$. Numerical integration shows stability of solitons at all
values of the charge and at both values of the transverse rigidity
$K=0.234$ N/m and $K=0.8714$ N/m. The viscosity of the environment
leads to quick stop of the soliton, and after that all time it
remains immovable. Soliton remains stable with respect to thermal
oscillations during the all time of numerical integration
$t=5\times 10^3$ ps (Fig. \ref{fig13}).

Let us note that in
contrast to the models of phi-4 and of sine-Gordon
the stability of solitons in the DNA model has not topological
nature. Solitons can be destroyed.
To show this, it is enough to suggest that the soliton width is equal
to one base pair (soliton of that type is equivalent to the ground
state of the chain). Here the stability
is associated with energetic factors. From Fig. \ref{fig13}
it is well seen that  in the region
of localization of the soliton, the density of the energy
is equal to $E_n\gg k_BT$.
%-------------------------------------------------------------------------
\begin{figure}
\centering\epsfig{file=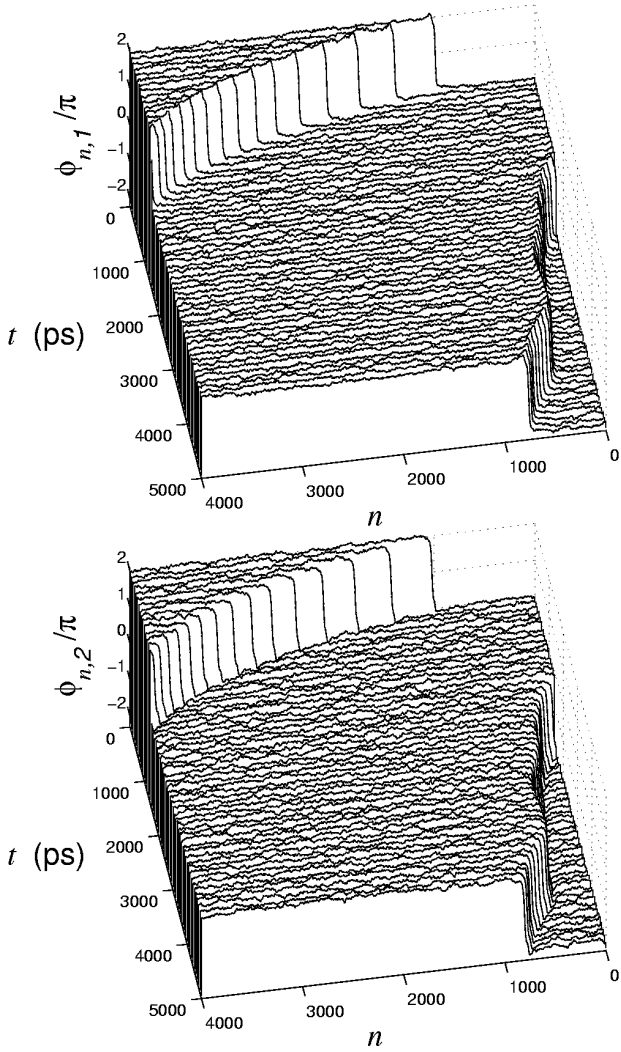,width=0.99\linewidth}
\caption{\label{fig14}\protect
        The braking of topological soliton (charge {\bf q}$=(1,1)_l$,
        initial velocity $s=0.5$) in the thermalized cyclic
        homogeneous AT chain ($T=300$K, $K=0.234$ N/m and
        $t_r=1000$ps).
        }
 \end{figure}
%-------------------------------------------------------------------------

Soliton path length in the thermalized homogeneous chain
($T=300$K) depends on the value of relaxation time $t_r$ (on the
viscosity of the surrounding of the molecule). At strong viscosity
$t_r=1$ ps soliton has time to pass only 7 chain links till full
stop. Then it remains immovable all the time (Fig. \ref{fig13}).
When the viscosity is lower $t_r=10$ ps soliton has time to pass
41 links, and when $t_r=100$ ps -- 480 links. The braking of
soliton at low viscosity ($t_r=1000$ ps) is shown in Fig.
\ref{fig14}. Soliton passes more than 3000 chain links, and then
it begins to move as a massive Brownian particle.

The braking of soliton in the homogeneous chain is conditioned
only by viscosity. When the viscosity is absent ($t_r=\infty$)
soliton is moving along thermalized chain with constant velocity
(Fig. \ref{fig15}).Thetmal phonons by themselves do not influence
the soliton dynamics.
%-------------------------------------------------------------------------
\begin{figure}
\centering\epsfig{file=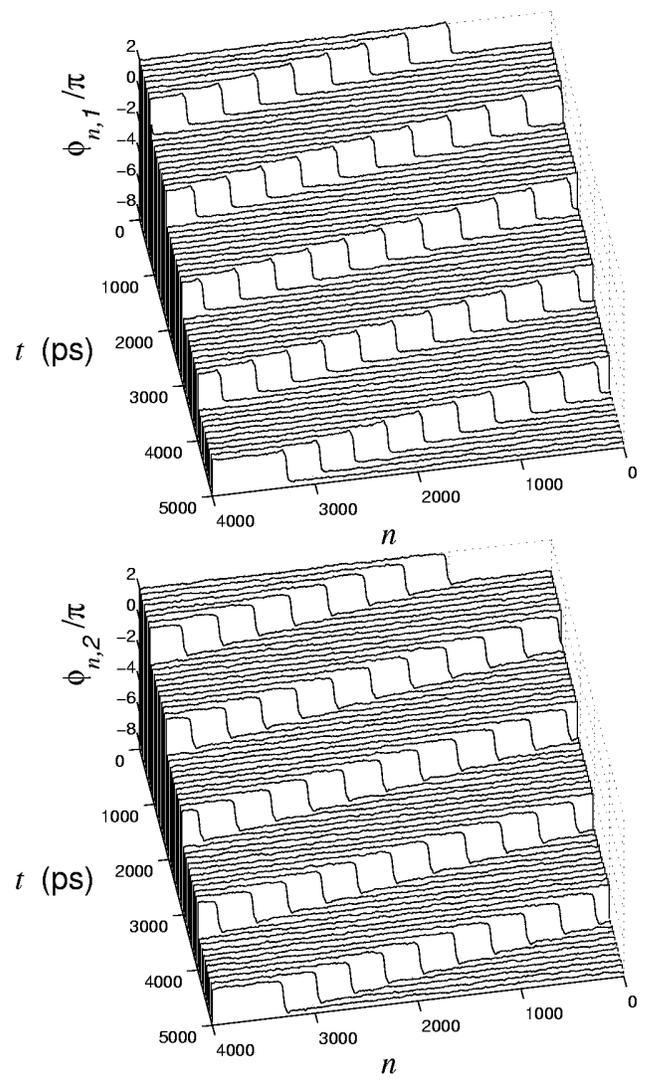,width=0.99\linewidth}
\caption{\label{fig15}\protect
        Movement of topological soliton (charge {\bf q}$=(1,1)_l$,
        initial velocity $s=0.5$) in the thermalized cyclic
        homogeneous AT chain ($T=300$K, $K=0.234$ N/m and
        $t_r=\infty$).
        }
 \end{figure}
%-------------------------------------------------------------------------
%-------------------------------------------------------------------------
\begin{figure}
\centering\epsfig{file=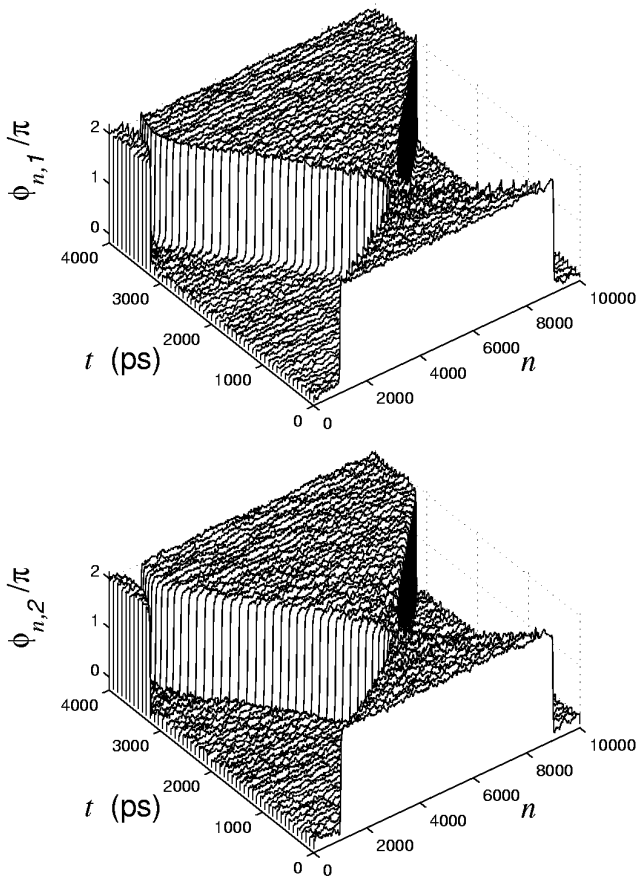,width=0.99\linewidth}
\caption{\label{fig16}\protect
        Reflection of topological solitons with different charges
        and polarities
        (${\bf q}_1=(1,1)_l$, ${\bf q}_2=(-1,1)_r$,
        $s_1=-s_2=0.5$) in the thermalized $(T=300$K) cyclic
        homogeneous AT chain ($K=0.234$ N/m, $t_r=\infty$).
        }
 \end{figure}
%-------------------------------------------------------------------------

Let us note that topological soliton can move along the DNA chain
in the presence of viscosity too. To organize the propagation it
is necessary to select in a special way the sequence of bases. If
concentration of AT base pairs monotonically increases, inclined
potential $E(n)$ is formed. The energy overfall can reach 1116
kJ/mol at $K=0.234$ N/m and 1894 kJ/mol at $K=0.8714$ N/m. Soliton
will propagate along the relief inclination as a Brownian particle
moving in the viscous media under the action of external constant
force.

Thermal phonons substantially influence the interaction of
solitons. In the work \cite{p42}, it was shown that topological
solitons of the model $\phi$--4, can interact with one another
through thermal phonons.
This interaction comes to
repulsion of the solitons. As a result, in the thermalized chain
the interaction of the solitons of different charges
substantially changes.
At a long distance they will repulse.
To model this phenomenon, let us consider collision
of solitons with different charges and polarities in
the thermalized cyclic AT chain
$({\bf q}_1=(1,1)_l$, ${\bf q}_2=(-1,-1)_r$,
$s_1=-s_2=0.5$). In the nonthermalized chain ($T=0$K), the solitons
attract one another, and the collision leads to their recombination.
In the thermalized cyclic chain ($T=300$K), their collision always
leads to reflection (Fig. \ref{fig16}). This behavior can be
explained by compression of phonons gas between solitons when they
are drawing together.  The compression leads
to the repulsion of solitons, which increases as far as they
are drawing together. In the chain with free ends, the compression
of the phonon gas leads to
long range repulsion of the solitons from the ends of the chain.

Thus, topological solitons of the DNA chains are stable with respect to
thermal oscillations. Interaction with thermal phonons does not
lead to destruction or to the braking of the soliton, it leads only to
changing the interaction between the solitons.
In the thermalized chain, long range repulsion between solitons
is appeared.

\section{Conclusion}
Investigation carried out in this paper shows that three types of
topological solitons which imitate localized states with open base
pairs, can exist in the considered asymmetrical model of the DNA double
chain. It was shown that the solitons can move along
the macromolecule with constant velocity which is smaller than the
sound velocity. In the inhomogeneous chain, the character of
the soliton movement depends on the sequence of base pairs in
the molecule. In the chain with random
inhomogeneous sequence, solitons can move at a distance no more than
several hundreds of base pairs. The results of numerical investigations
show that the solitons are stable with respect
to thermal oscillations. Interaction of the solitons with thermal
phonons of the macromolecule does not lead to destruction or to
the braking of the solitons. And only the character of their
interactions changes. The drawing of the solitons together
leads to their repulsion, which is explained by compression of
phonon gas between them.

All these results point out that topological solitons of this
type can be used to explain the long range effects in the DNA
macromolecule.

\end{document}